\documentclass[12pt]{article}
\usepackage[utf8x]{inputenc}

\usepackage{amsfonts,amssymb}
\usepackage{amsmath}
\usepackage{bm,graphics}

\newcommand{\R}{\mathbb{R}}

\newcommand{\I}{\mathrm{i}}

\newcommand{\ex}{\mathbf{E}}

\newcommand{\ffield}{A}
\newcommand{\LL}{\mathcal{L}}

\newcommand{\yzero}{y_0}
\newcommand{\siminf}{\bumpeq}
\newcommand{\ma}{\bm{a}}
\newcommand{\mb}{\bm{b}}
\newcommand{\ml}{\bm{a}}
\newcommand{\mq}{\bm{q}}
\newcommand{\mI}{\bm{I}}
\newcommand{\mtxu}{\bm{u}}
\newcommand{\mtheta}{\bm{\theta}}
\newcommand{\msigma}{\bm{\sigma}}
\newcommand{\deltahat}{\hat{\delta}}
\newcommand{\gammahat}{\hat{\gamma}}
\newcommand{\fy}{f}
\newcommand{\gy}{g}
\newcommand{\hy}{h}
\newcommand{\ny}{n}
\newcommand{\muinfty}{\mu_\infty}
\newcommand{\spiel}{(a) PDE solver, (b) leading-order approximation, (c) comparison at several different time points (dotted line is PDE solver, solid is leading-order approximation).}
\newcommand{\sbkt}[1]{{\textstyle(}#1{\textstyle)}}

\newcommand{\notthis}[1]{}

\newcommand{\inv}{^{-1}}

\newcommand{\half}{\frac{1}{2}}

\newcommand{\cdl}{\,|\,}

\newcommand{\shalf}{{\textstyle\frac{1}{2}}}

\newcommand{\pderiv}[2]{\frac{\partial{#1}}{\partial{#2}}}
\newcommand{\deriv}[2]{\frac{d{#1}}{d{#2}}}

\newcommand{\pdderiv}[2]{\frac{\partial^2{#1}}{\partial{#2}^2}}
\newcommand{\sech}{\mathrm{sech}}

\newcommand{\Beta}{\mathrm{B}}
\newcommand{\tr}{\mathrm{tr}}

\begin{document}

\title{\bf Analytical approximation to the multidimensional Fokker--Planck equation with steady state}
\author{R. J. Martin\footnote{Department of Mathematics, Imperial College London, London SW7 2AZ, UK }\mbox{ }, R. V. Craster$^*$, A. Pannier$^*$ and M. J. Kearney\footnote{Senate House, University of Surrey, Guildford, GU2 7XH, UK }
}


\maketitle 

\begin{abstract}


The Fokker--Planck equation is a key ingredient of many models in physics, and related subjects, and arises in a diverse array of settings. Analytical solutions are limited to special cases, and resorting to numerical simulation is often the only route available; in high dimensions, or for parametric studies, this can become unwieldy. Using asymptotic techniques, that draw upon the known Ornstein--Uhlenbeck (OU) case, we consider a mean-reverting system and obtain its representation as a product of terms, representing short-term, long-term, and medium-term behaviour. A further reduction yields a simple explicit formula, both intuitive in terms of its physical origin and fast to evaluate. We  illustrate a breadth of cases, some of which are `far' from the OU model, such as double-well potentials, and even then, perhaps surprisingly, the approximation still gives very good results when compared with numerical simulations. Both one- and two-dimensional examples are considered.

Published in J.Phys.A (Math.Theor.), {\bf 52}(8):085002 (2019)
\end{abstract}



\section{Introduction}

The Fokker--Planck equation (FPE) arises in a broad range of problems from physics, engineering science, economics and mathematical modelling. Part of this breadth of application can be traced back to its success in modelling generic transport processes provided the dynamics can be represented by a Hamiltonian or Lagrangian with random components \cite{escande07a}. Alternatively, a time series of data can be analysed as a Markov process to create an effective FPE that captures the statistics of the observed process. This has led to modelling based upon extracting FPEs from experimental or observable data for instance in turbulent cascades \cite{friedrich97a}, fractal-generated turbulence \cite{stresing10a}, modelling the beat fluctuations in heart-rate \cite{ghasemi06a}, electronic noise and kinetics \cite{Coffey03}, electronic circuits with nonlinear resistance \cite{Dubkov09}, 
systems with overdamped Langevin dynamics \cite{Dybiec07a},
or from nonlinear friction \cite{Klimontovich94,Touchette10a}.  
Financial modelling yields a wealth of further applications \cite{friedrich00a}. The
modelling of market behaviour, where deviation from equilibrium is likely to be accompanied by higher volatility and/or slower mean-reversion, means that the invariant distribution is fat-tailed or leptokurtic \cite{Martin15b}.
Agent-based models \cite{alfarano05a} for the herd behaviour of interactions between traders, and the influence of rumours, social interactions and external information \cite{carro15a}, take ideas from Kirman's stochastic models of information transmission \cite{kirman93a} to arrive at FPEs.
Rather than study a single specialised case we consider generic FPEs with the main restriction being that we consider mean reverting processes. 

Another important application for FPEs is in Kalman filtering, which in addition to physics also impacts upon control theory, optimisation, and time series analysis. Examples include: 
 subatomic particle tracks \cite{Cervera02}; movements in the ionosphere \cite{Scherliess04}; chemical reactions \cite{Ruess15}; and extensive use in econometrics for making predictions about financial variables in systems dominated by stochastic behaviour, e.g.~\cite{Doz11}.
The basic operation of Kalman-type filters is long-established (see e.g.~\cite{maybeck79a,harvey89a,grewal93a}), and a key ingredient is the state transition density. Initial work used linear assumptions, i.e.\ an underlying Gaussian model. When the drift and covariance are nonlinear functions of the state vector, the extended Kalman filter \cite{einicke99a} is often used, but in essence this employs a linearisation so that the Gaussian is used locally. This is usually acceptable for short time periods, but may not be over longer ones. A better approximation to the transition density for non-Gaussian processes, such as that we provide, is therefore highly desirable.

The normalised form of the FPE, in one dimension, that we investigate is
\begin{equation}
\pderiv{f}{\tau} =  -\pderiv{}{y} \big[ \ffield(y) f \big] + \pdderiv{f}{y} \equiv \LL^\dagger f ,
\qquad
f(0,y)=\delta(y-\yzero)
\label{eq:pde_fy}
\end{equation}
 where $f$ is the probability density function, $\tau$ is nondimensional time and $y$ is a spatial variable; the initial condition is a delta-function centred at $\yzero$. There is a general drift function $\ffield(y)$ for which we will take various choices as examples in later sections. The steady-state solution $f(\infty,y)\propto\exp\int^y \ffield(z)\,dz$, is a normalisable probability density: this can be ensured by the condition $\limsup_{|y|\to\infty} -y \ffield(y) > 1+\varepsilon$ for some $\varepsilon>0$.
As alluded to earlier this partial differential equation, the FPE, is connected to a stochastic differential equation (SDE):
\begin{equation}
dY_t = \kappa \ffield(Y_t) \, dt + \sqrt{2\kappa} \, dW_t ,
\label{eq:sde_y}
\end{equation}
with $\tau=\kappa t$. 
More generally an SDE having both spatially varying mean and variance is
\begin{equation}
dX_t = \mu_X(X_t) \, dt + \sigma_X(X_t) \, dW_t;
\label{eq:sde_x}
\end{equation}
however, provided $\sigma_X$ is bounded away from zero we can make the substitution (also known as the Lamperti transformation) from $X$ to $Y$ defined by $dy/dx=\sqrt{2\kappa}/\sigma_X(x)$, which places (\ref{eq:sde_x}) in the normalised form (\ref{eq:sde_y}); therefore the analysis we present for the normalised form is more general than it first appears. In higher dimensions we take $\ffield$ to be the gradient of a potential, i.e.\ conservative, and note that under some common assumptions the Lamperti transformation can be appropriately generalised \cite{moller10a}. It is worth noting that in one particular case, the \emph{affine} model in which $\mu$ and $\sigma^2$ are both linear in $X_t$, the FPE associated with \eqref{eq:sde_x} is exactly solvable \cite{gihman72a,lefever74a,Lamberton12}. In general, though, the full solution of the FPE is considerably more difficult than the stationary solution.

The best-known example with a steady state is the OU process given by $\ffield(y)=-\theta y$ ($\theta$ is a parameter), for which there is a well-known explicit Gaussian solution \cite{risken89a}. However, one often wants to deviate from this model because away from equilibrium the force field $\ffield$ cannot be expected to rise without limit, but instead be bounded; equivalently, the equation describes diffusion in the presence of a potential which cannot be expected to be quadratic in general. In these situations the steady state will no longer be Gaussian. Departing from the OU, whilst gaining closer connection to the physical model under consideration, loses analytical tractability and typically numerical methods are required. Naturally, one would like the best of both worlds: analytical tractability and physical relevance. 

Given the connection with SDEs, there is a choice between the deterministic or stochastic approaches in terms of which is more practical to tackle numerically. Choosing the latter naturally leads to Monte-Carlo methods: evaluating the density of a stochastic process requires not only a large number of simulations, but also kernel density estimation (KDE) at each point in time to produce a smooth estimate from the simulated data points: For literature on KDE, see. e.g.\ \cite{Scott92,Silverman86,Buch05}.
The quality of the estimation depends on the kernel width and the number of simulations. The optimal choice of width is not straightforward \cite{Devroye85,Jones96}\footnote{The bandwidth that minimizes the mean integrated square error, Scott's rule, is of order $N^{-1/(d+4)}$ where $N$ is the number of simulations and $d$ the number of dimensions and leads to an accuracy of $O(N^{-4/(d+4)})$}. In the work we have done, if $\fy(\infty,y)$ is very fat-tailed, for example Student-t, then simulation of $\gtrsim 100,000$ paths is required to get 
an $L^1$ error (integrated absolute error) of $4.10^{-2}$. In two dimensions, $\gtrsim 100,000$ paths only yield
an $L^1$ error of $2.10^{-1}$.
Furthermore, for every evaluation point one typically needs to make use of every path simulated, so the computation time of estimating the density is much higher than the simulation time and suffers considerably from the curse of dimensionality; see \cite{chen18a} for background and numerical algorithms that aim to lift the curse. This impediment means that numerical methods for the FPE are preferred and literature combining Monte-Carlo and KDE methods is rare and specific, e.g.~\cite{Burke16}. Thus, when we perform numerical calculations we do so upon the FPE directly; for efficiency and accuracy we use spectral methods and detail these in section \ref{sec:1dexamples}. Even so, the numerical simulations in two dimensions become very time-consuming, particularly if one is not looking for the solution at short time. Furthermore, if one desires to sweep over different parameter values to analyse the effect on the results, the problem gets even worse.

All this points to the need for a simple, fast, approximation. Further, it is desirable to have a modus operandi that gives results that are intuitive and offer direct insight into the problem at hand.
Ideally, we would like to address the metaphysical question of `how diffusions think about solving themselves'.
In this respect, what we are going to describe---in the first instance (\ref{eq:newf1})---does give clear intuition, in that the various terms in the equation make it clear how the solution behaves; also, the result at some basic level \emph{looks} like an expression for the evolution of a probability density, in a way that an infinite series of eigenfunctions does not.

Most analytical approximations are based on the summation of eigenfunctions \cite{risken89a} that are often orthogonal polynomials and special functions, which is unsurprising given the linearity of the FPE;  the OU can also be approached this way using Hermite polynomials \cite{Mehler66}. Alternative asymptotic approaches, such as WKB methods \cite{caroli79a}, are useful for analysing the approach to equilibrium but are limited to studying specific regimes involving small or large diffusivities. 
 A somewhat different approach, upon which the first steps were made in \cite{Martin15b}, consists in expressing the solution as a product of terms, rather than the more usual sum.
Intuitively, with a sum it is difficult to represent the initial delta-function without generating oscillatory artefacts, whereas with a product this is simple: a narrow Gaussian, of width tending to zero as $\tau\to0$, will capture that \emph{regardless of what the other terms in the product are}; another term can capture the steady state; and a series of correction terms `patches-up' the mid-term behaviour. Intuitively, we are expanding around an OU model, in the sense of finding the characteristics of a mean-reverting solution as exemplified by the OU case and capturing these characteristics for the general case, while reproducing the OU case exactly.
There is a passing similarity with the WKB approximation, mainly in the use of a logarithmic transformation, but WKB expands around a zero-volatility (deterministic) model and is singular in the zero-volatility limit.

There are some important consequences of using products.
The logarithm of the density is represented by a sum, and so: (i) positivity is guaranteed, in a way that it is typically not using linear methods; (ii) from the theoretical perspective there is a clear relation to entropy, and calculation of that from an approximated density is virtually impossible if the approximated density is anywhere negative. Based on these points, we therefore choose to develop this approach.

We begin by introducing our approach in the one-dimensional setting (Section~\ref{sec:oneD}), giving the key results and the main technical route to them. A range of examples demonstrates the efficiency of the results as we move away from the OU process. We then move on to the higher-dimensional case in Section~\ref{sec:multi} and show numerical simulations. Section~\ref{sec:extensions} discusses extensions and potential limitations of the method, with concluding remarks drawn together in Section~\ref{sec:conclude}.

\section{Theory in one dimension}
\label{sec:oneD}
We introduce the methodology and results in one dimension. A key step is the decision to work with the normalised density $g$, and the derivative, $h$, of the log-density  
\footnote{$f$, $g$, $h$ are understood to mean $f_Y$, $g_Y$, $h_Y$ respectively.} defined as $\hy = -(\partial/\partial y) \log \gy$, with $\gy(\tau,y)=\fy(\tau,y)/\fy(\infty,y)$. Thereby $\gy$ solves the backward equation
\begin{equation}
\pderiv{\gy}{\tau} = \ffield(y) \pderiv{\gy}{y} + \pdderiv{\gy}{y} \equiv \LL g
\label{eq:pde_g}
\end{equation}
with initial condition a delta-function of strength $1/\fy(\infty,\yzero)$ at $\yzero$, and $\hy$ solves the nonlinear partial differential equation (PDE)
\begin{equation}
\pderiv{\hy}{\tau} = \pderiv{}{y} \left\{ 
\ffield(y) \,  \hy +  \pderiv{\hy}{y}-\hy^2  
\right\}
\label{eq:pde_h}
\end{equation}
which has a singular initial condition, in the sense that $\hy \sim (y-\yzero)/2\tau$ as $\tau\to0$.
This singular behaviour can also be seen by dominant balance in (\ref{eq:pde_h}).

Next we observe that for the OU model $\ffield(y)=\theta(y_\infty-y)$, with the constant $y_\infty$ denoting the long-term mean, we have exactly
\begin{equation}
\mbox{(OU)}\qquad \hy(\tau,y) = \frac{\theta \!\sqrt{q} (y-\yzero)}{1-q} + \frac{\theta\!\sqrt{q}(y_\infty-y)}{1+\!\sqrt{q}} , \qquad q=e^{-2\theta \tau},
\label{eq:OU}
   \end{equation}
as is easily verified by substituting it into (\ref{eq:pde_h}) and ploughing through the algebra.
Finally, from its definition, $\gy(\infty,y)$ must always equal unity, and so $\hy(\infty,y)=0$. This inspires the ansatz 
\begin{equation}
\hy(\tau,y) = \frac{\theta \!\sqrt{q} (y-\yzero)}{1-q} + \frac{\sqrt{q}}{1+\!\sqrt{q}} \ffield(y) + \sqrt{q}\,o(1)_{q=1} 
\label{eq:hynew}
\end{equation}
for arbitrary $\ffield$.
When (\ref{eq:hynew}) is inserted into the PDE for $h$, (\ref{eq:pde_h}), and a Laurent expansion performed around $\tau=0$, the LHS and RHS agree at $O(\tau^{-2})$ and $O(\tau^{-1})$, explaining why we are writing the error term in (\ref{eq:hynew}) as $o(1)$ in the short-time limit.
This error can then in principle be approximated as a Taylor series around $\tau=0$, which we will discuss later. 

Another matter presents itself when this Laurent expansion is done: the development of (\ref{eq:hynew}) is 
\[
\hy(\tau,y) = \frac{y-\yzero}{2\tau} + \frac{\ffield(y)}{2} + o(1), \qquad \tau\to0,
\]
and we observe that $\theta$ is absent from both the first two terms.
Accordingly, all $\theta$'s give the same leading-order behaviour, and so we cannot say anything about $\theta$ simply by looking at the first two terms in the short-time expansion of the solution. In this sense, therefore, $\theta$ is now arbitrary, representing an estimate of the mean reversion speed of the force-field $\ffield$, or, equivalently, the reversion speed of the OU process `about which' we are expanding the given model $\ffield$. Given that the leading-order asymptotics do not tell us what $\theta$ to use, some other method of inference is necessary, and we return to this later. 
In this context we call \eqref{eq:hynew} the leading-order approximation for $\hy$.

To deduce $g$ and $f$ we integrate (\ref{eq:hynew}) from $\yzero$ to $y$, giving (again for arbitrary $\ffield$)
\[
\gy(\tau,y) \sim (\ldots)_{\tau,\yzero} \exp\left( 
\frac{-\half\theta \!\sqrt{q}(y-\yzero)^2}{1-q} \right)
\fy(\infty,y)^\frac{\scriptstyle -\sqrt{q}}{\scriptstyle 1+\sqrt{q}}
\fy(\infty,\yzero)^\frac{\scriptstyle \sqrt{q}}{\scriptstyle 1+\sqrt{q}}
\]
where $(\ldots)_{\tau,\yzero}$ generically denotes a function of $\tau,\yzero$. 
By means of the reciprocity condition \footnote{This can also be written ${\fy(\tau,y \cdl \yzero)}/{\fy(\infty,y)} = {\fy(\tau,\yzero \cdl y)}/{\fy(\infty,\yzero)}$. Viewed as a function of $y$ (and $\tau$), the LHS obeys the adjoint forward equation, whereas the RHS obeys the backward equation. However, those are the same PDE, with the same initial condition; alternatively, we could invoke the Kolmogorov criterion, e.g.\ \cite[\S1.5]{Kelly79}.}
\begin{equation}
\gy(\tau,y \cdl \yzero) = \gy(\tau,\yzero \cdl y)  
\label{eq:revers}
\end{equation}
that is obeyed by the exact solution, we can infer the dependence of the prefactor on $\yzero$, to obtain
\[
\gy(\tau,y) \sim (\ldots)_{\tau} \exp\left( 
\frac{-\half\theta \!\sqrt{q}(y-\yzero)^2}{1-q} \right)
\fy(\infty,y)^\frac{\scriptstyle -\sqrt{q}}{\scriptstyle 1+\sqrt{q}}
\fy(\infty,\yzero)^\frac{\scriptstyle -\sqrt{q}}{\scriptstyle 1+\sqrt{q}}
.
\]
The OU case requires the prefactor to be
\[
\frac{(\theta/2\pi)^\frac{\scriptstyle \sqrt{q}}{\scriptstyle 1+\sqrt{q}}}{\sqrt{1-q}}. 
\]
(Another way of deriving the prefactor is to write
\[
\gy(\tau,y) = \ny(\tau) e^{-\int_{\yzero}^y \hy(\tau,z) \, dz},
\]
substitute into (\ref{eq:pde_fy}), and solve for the function $\ny$, which obeys a first-order linear differential equation; we shall refer to this technique later on.)  
Thence 
\begin{equation}
\gy(\tau,y) \sim  \frac{1}{\sqrt{1-q}} 
\exp\left( 
\frac{-\half\theta \!\sqrt{q}(y-\yzero)^2}{1-q}    
\right)
\left(\frac{\theta/2\pi}{\fy(\infty,y) \fy(\infty,\yzero)} \right)^\frac{\scriptstyle \sqrt{q}}{\scriptstyle 1+\sqrt{q}}
\label{eq:newg1}
\end{equation}
and 
\begin{equation}
\fy(\tau,y) \sim  \frac{(\theta/2\pi)^\frac{\scriptstyle \sqrt{q}}{\scriptstyle 1+\sqrt{q}}}{\sqrt{1-q}} 
\exp\left( 
\frac{-\half\theta \!\sqrt{q}(y-\yzero)^2}{1-q}    
\right)
\fy(\infty,y)^\frac{\scriptstyle 1}{\scriptstyle 1+\sqrt{q}} \fy(\infty,\yzero)^\frac{\scriptstyle -\sqrt{q}}{\scriptstyle 1+\sqrt{q}} 
\label{eq:newf1}
\end{equation}
which are the lowest-order approximations. 

Eq.~(\ref{eq:hynew}), and its consequences, have several facets worthy of comment. First, they are exact for any OU model of reversion speed $\theta$ regardless of the reversion level, i.e.\ for $\ffield(y)=\theta (y_\infty-y)$.
Another way of putting this is to say that the correction terms to (\ref{eq:hynew}) will be expressible as functions of $(d/dy)(\ffield(y)+\theta y)$.
Secondly, the approximations (\ref{eq:newg1},\ref{eq:newf1}) are necessarily positive, and  correct in both the short- and long- time limits, regardless of $\ffield(y)$.
Thirdly, the approximations (\ref{eq:newg1},\ref{eq:newf1}) obey the reciprocity condition.

We can proceed to determine higher-order terms in the representations for, say, $h$ by 
  continuing (\ref{eq:hynew}) and introducing an expansion for the remainder so
\begin{equation}
\hy(\tau,y) = \frac{p}{1+p} \left( \frac{\theta (y-\yzero)}{1-p} + \ffield(y) + \sum_{r=1}^\infty (1-p)^r b_r(y) \right).
\label{eq:hynew2}
\end{equation}
 where $p=\sqrt{q}=e^{-\theta \tau}$, and consider the remainder terms, $(b_r)$, that also depend parametrically on the starting-point $\yzero$. An alternative expansion is in powers of $(1-q)$, giving a different series, but with similar convergence properties and we do not pursue this further here. The power of $\sqrt{q}$ in \eqref{eq:hynew2} ensures $h(\infty,y)=0$ for any truncation of the series (i.e.\ sum as far as $r=N$).
By inserting (\ref{eq:hynew2}) into (\ref{eq:pde_h}) and comparing coefficients in powers of $(1-p)^{r+1}$ we find that
\[
\deriv{}{y} \big[(y-\yzero)^{r+2}b_{r+1}(y)\big] = (y-\yzero)^{r+1} \mathfrak{F}_r(y) 
\]
where $\mathfrak{F}_r$ is a complicated quadratic expression invoking $\ffield,b_1,\ldots, b_r$ and their derivatives.
So although (\ref{eq:pde_h}) is second-order nonlinear, successive terms in the expansion can be recursively obtained by solving a first-order linear differential equation, which can be integrated immediately to give\footnote{The lower limit has to be $\yzero$, as otherwise $b_{r+1}$ will be singular at $y=\yzero$.}
\[
b_{r+1}(y) = \frac{1}{(y-\yzero)^{r+2}} \int_{\yzero}^y (z-\yzero)^{r+1} \mathfrak{F}_r(z)  \, dz .
\]
In the special case $r=0$, we have
\[
b_1(y) = \frac{1}{\theta(y-\yzero)^2} \int_{\yzero}^y (z-\yzero) \left( \deriv{}{z} + \ffield(z) + \frac{\theta}{2}(z-\yzero) \right) \deriv{}{z} \big(\ffield(z)+\theta z\big) \, dz
\]
which vanishes whenever $\ffield(y)+\theta y$ is a constant, as it should. 
As an aside, the term $b_1(y)$ also vanishes in another special case, i.e. when 
\[
\widetilde{\ffield}' + \shalf \widetilde{\ffield}^2 = \theta (y^2/4 + \yzero/2 + \mbox{const}), \qquad \widetilde{\ffield}(y) \equiv \ffield(y)+\theta y;
\]
which has $\widetilde{\ffield}$ as the logarithmic derivative of a parabolic cylinder function. This solution is sporadic in that it also depends on the starting-point, that is, for this choice of $\ffield(y)$ the function $b_1=0$ only if $\yzero$ is chosen correctly.

As (\ref{eq:hynew}) is a sum, the expressions for $\fy,\gy$ will be infinite products, and so we refer to the method as an infinite product expansion. The focus here is not on the extraction of higher-order terms: rather, it is on the leading-order term, which is probably the most applicable. But it is notable that, should it be desired, one can also treat the correction term by spectral methods: that is to say, write $\widetilde{f}(\tau,y)$ for (\ref{eq:newf1}), derive the PDE that it satisfies (this is another parabolic PDE), and solve it approximately by means of a Galerkin or collocation expansion \cite{Boyd01}. As the initial spike and also the long-term behaviour have already been accounted for in $\widetilde{f}$, the unknown function $f/\widetilde{f}$ is unity in both limits $\tau\to0,\infty$, and therefore well approximated by spectral methods, which are ideally suited to smooth problems. 

The first steps towards a product expansion were made in \cite{Martin15b}, which carries out a power series expansion on broadly similar lines and provides an expansion of the form
\begin{equation}
\hy(\tau,y) = \frac{\theta (qy- \!\sqrt{q}\,\yzero)}{1-q}  + \sqrt{q} \sum_{r=0}^\infty (1-q)^r b^\textrm{old}_r(y).
\label{eq:hyold}
\end{equation}
The development here confers several advantages, besides greater compactness and elegance, over \cite{Martin15b}: (i) the reciprocity condition (\ref{eq:revers}) is enforced, whereas it was not in \cite{Martin15b}; (ii) it is more readily adaptable to higher dimensions; (iii) it is more accurate over shorter time-scales; (iv) the leading-order approximation in \cite{Martin15b} exhibits instability when $|\yzero|$ is large.
What is shown here can be obtained from (\ref{eq:hyold}) by taking the initial term and $b^\textrm{old}_0(y)$ as the new leading-order term, with minor alterations, and modifying the prefactor in the derivation of $\gy$ so as to enforce (\ref{eq:revers}). Otherwise, however, it is neither more nor less convergent, being in effect a rearrangement of the terms.

The approach of expanding around an OU process leaves a free parameter, $\theta$, that controls the intermediate-time behaviour, and the remainder series implicitly depends upon it. As we said earlier, we cannot infer $\theta$ from the $O(\tau^{-1})$ or $O(\tau^{0})$ terms in the short-time expansion. We can argue that  $\theta$ should be chosen to minimise $b_1(y)$ as given earlier, and as that function vanishes when $\ffield'(y)+\theta$ is identically zero, it seems reasonable to choose $\theta$ so as to minimise $\ffield'(y)+\theta$ `on average'. This motivates the choice  \begin{equation}
\hat{\theta} = \langle -\ffield' \rangle_\infty = \langle \ffield^2 \rangle_\infty
\label{eq:theta}
\end{equation}
where $\langle \cdot \rangle_\infty$ means an average over the invariant density $\fy(\infty,\cdot)$.
This was proposed in \cite{Martin15b}, albeit using a different line of reasoning based on a sort of Rayleigh-Ritz argument to identify the least negative eigenvalue of $\LL^\dagger$, the differential operator of the FPE\footnote{It was, however, reliant on the operator $\LL^\dagger$ having a discrete spectrum, which it may not.}.
Clearly  (\ref{eq:theta})  is correct in the OU case, and it guarantees $\hat{\theta}>0$, all of which make it a pragmatic choice, but there is another compelling reason based on a connection with entropy and information theory, which runs as follows.
Consider, for some p.d.f.~$\psi$, the family of distributions $\psi(y-\mu)$ indexed by the parameter $\mu\in\R$. It is desired to estimate $\mu$ (from data), and the standard way of doing this is the maximum likelihood estimator. Writing
\[
f(y \cdl \mu) = \psi(y-\mu)
\]
we seek to maximise $\log f(y \cdl \mu)$ w.r.t.\ $\mu$. The \emph{Fisher information} \cite[\S2.5]{Lehmann98} is the expectation of the square of the $\mu$-derivative of the log-likelihood, and hence is 
\begin{eqnarray}
\int_{-\infty}^\infty \left(\pderiv{}{\mu} \log f(y\cdl\mu)\right)^2 f(y\cdl\mu) \, dy \nonumber
&=&
\int_{-\infty}^\infty \left(\frac{\psi'(y-\mu)}{\psi(y-\mu)} \right)^2 \psi(y-\mu) \, dy \\
&=& \int_{-\infty}^\infty \left(\frac{\psi'(y)}{\psi(y)} \right)^2 \psi(y) \, dy; \nonumber
\end{eqnarray}
if we set $\psi(y)=\fy(\infty,y)$ then this is exactly the definition of $\hat{\theta}$.
In broad terms, the higher the Fisher information, the more certain we are about the estimation of the parameter in question, and indeed the reciprocal of the Fisher information furnishes the Cram\'er-Rao lower bound for the variance of any unbiased estimator. This has an interpretation in terms of the mean reversion: 
 the higher the average speed of mean reversion, the more certain we are about our estimate of the mean from a given dataset, and vice-versa. Using the Fisher information as an estimator of reversion speed is therefore  natural. As will be seen later, the method extends in a natural way to higher dimensions, with $\partial/\partial\mu$ replaced by $\nabla$, and then the Fisher information is a positive definite symmetric matrix rather than just a positive number.


\subsection{Results and discussion in one dimension}
\label{sec:1dexamples}


We consider a range of one dimensional examples deviating from the OU by different degrees and compare the leading-order approximation (\ref{eq:newf1}) to numerical solutions from a PDE solver. The numerical solution is computed by means of Fourier spectral collocation in the spatial direction coupled with a fourth-order Runge-Kutta algorithm in time and uses a narrow Gaussian as initial condition; a sufficiently large spatial domain is taken such that the FPE does not interact with the edge, and convergence is checked by mode-doubling. We take advantage of the linear diffusion by using an integrating-factor, and also the Fast Fourier Transform, to design a highly efficient solver; such methods are standard in scientific computing \cite{trefethen00,canuto06} and the fully-converged numerical simulations act as the `gold standard' against which we compare the approximations. Even then, the PDE solver is far slower than the leading-order approximations. Our solution is a truncation from an infinite product, and so it is of central importance to demonstrate its efficiency. 
The following examples explore the domain of validity of the leading-order approximation in which it successfully replicates the time-evolution of the numerical FPE solutions. We mainly choose cases from fat-tailed invariant distributions, which were our main motivation; the last case exhibits the adaptability of the approximation in places where it was unexpected. 

\subsubsection{Sech-power model}

One way of moving away from the linear force field (quadratic potential
well) of the OU model is to make the force field grow less rapidly
away from equilibrium by stipulating
\begin{equation}
\ffield(y)=-\frac{\deltahat}{\gammahat}\tanh\gammahat y.
\end{equation}
This is an example, of the well-known Pearson diffusions \cite{forman08}, which is obtained from the local volatility model
\begin{equation}
dX_{t}=-\kappa X_{t}\,dt+\sigma\sqrt{1+\gamma^{2}X_{t}^{2}}\,dW_{t},
\end{equation}
in which volatility increases away from equilibrium, using the transformation $\gamma X=\sinh\gammahat Y$; see \cite{Martin15b} for more details and some applications in mathematical finance. 
The associated potential is also known in mathematical physics as the P\"{o}schl-Teller potential \cite{flugge71} in the special case $\deltahat/\gammahat^2 = 2$. The resulting Schr\"{o}dinger equation is solvable in terms of special functions and its link with the FPE has been used to derive analytical solutions, see \cite{Brics13}. 
 The steady state is a sech-power: 
\begin{equation}
\fy(\infty,y)=\frac{\gammahat(\cosh\gammahat y)^{-\deltahat/\gammahat^{2}}}{\Beta\big(\frac{\deltahat}{2\gammahat^{2}},\half\big)},\qquad\langle-\ffield'\rangle_{\infty}=\frac{\deltahat^{2}}{\deltahat+\gammahat^{2}}
\label{eq:sech}
\end{equation}
with B denoting the Beta function; the explicit solutions mean that this is an attractive model that has been well-studied, for instance, for systems with nonlinear random vibrations   (\ref{eq:sech}) occurs, see \cite{Liu69}. 

In the limit where $\gammahat\to0$ we recover the OU model, so $\gammahat$ is a measure of the deviation from OU. We have looked at many parameter sets, the comparison of the leading order solution to full numerical simulations is consistently qualitatively pleasing, and a typical example in Figure~\ref{fig:4} shows the comparison.









\subsubsection{Dry-friction}

Dry-friction \cite{Touchette10a} is the limit of the sech-power model obtained when $\deltahat=\gammahat\to\infty$, and we then have a discontinuous $A(y)$ as 
\[
\ffield(y)=-\mathrm{sgn}\,y,\qquad \fy(\infty,y)=\frac{e^{-|y|}}{2},\qquad\langle-\ffield'\rangle_{\infty}=1.
\]
This case is interesting in terms of the physics it describes, but also as additionally the transition density is available in closed form \cite{Touchette10a}, e.g.~by the usual route of Laplace transforming the Fokker--Planck equation:
\begin{equation}
\fy(\tau,y \cdl \yzero) = \frac{e^{-(y-\yzero)^2/4\tau}}{\sqrt{4\pi \tau}} e^{-\tau/4} e^{(|\yzero|-|y|)/2}
+ \frac{e^{-|y|}}{2} \Phi\!\left( \frac{\tau - |y| - |\yzero|}{\sqrt{2\tau}} \right)
\label{eq:f_dryfric}
\end{equation}
or
\begin{equation}
\gy(\tau,y \cdl \yzero) = \frac{e^{-(y-\yzero)^2/4\tau}}{\sqrt{\pi \tau}} e^{-\tau/4} e^{(|\yzero|+|y|)/2}
+ \Phi\!\left( \frac{\tau - |y| - |\yzero|}{\sqrt{2\tau}} \right) 
\label{eq:g_dryfric}
\end{equation}
 and this exact solution provides a convenient benchmark against which to test our theory.

In this example, $\fy$ arises as a sum of two pieces, and so $\hy$, rather than being a simpler function than $\gy$ (as it is for example with the OU), is more complicated. Nonetheless, comparing to numerical simulation the accuracy is excellent when starting at the origin (not shown) or fairly near the origin ($\yzero=-2$, Figure~\ref{fig:5}a), but worse when it is much further away ($\yzero=-5$, Figure~\ref{fig:5A}b). This points to a separate development of the theory that deals with far-field expansions, and that we pursue later in section \ref{sec:extensions}: as we will show by consideration of (\ref{eq:pde_h}), the first term in either of the above expressions corresponds to an approximation in which we start a long way from equilibrium and the drift is small, as here.

\notthis{
Of interest: (i) The spectrum is continuous in this case. (ii) Not only is $\fy$ essentially singular in $\tau$ at the origin, which is expected, but so too is $\hy$. Therefore $h$ has no Laurent expansion and taking more terms in the infinite product will not work (already anticipated in PRSA paper). This lends credence to the hypothesis that $\ffield(y)$ needs to be analytic. It doesn't mean that some other sort of infinite product can't work: it's just that you can't do an analytic expansion in $\tau$.
(iii) Essentially for short time the solution has to be understood in two segments: for $|y| + |\yzero|<\tau$ (a V-shaped domain) it behaves one way, and outside the V it behaves another way.
} 

\subsubsection{Student-t} 
\label{sec:student}
Another popular deviation, and as noted in the introduction important across many fields, from the OU is that associated with fat-tailed distributions. We use, as in \cite{Martin15b}, a distribution that conveniently has Student-t as its steady state in the $Y$ coordinates: 
\[
\ffield(y) = -\frac{y}{1+\gammahat^2 y^2} , \qquad
\fy(\infty,y) = \frac{\gammahat (1+\gammahat^2 y^2)^{-(\nu-1)/2}}{\Beta(\frac{\nu-2}{2},\half)}, 
\qquad 
\langle-\ffield'\rangle_{\infty}=\frac{\nu-2}{\nu+1}.
\]
This model gives rise to fatter tails than the sech-power example, because the force-field decays to zero as $|y|\to\infty$. The distinction is further accentuated by noting that the sech-power case can be obtained, for certain parameter values, from transforming a model in which the steady state is Student-t in $X$ coordinates; we demonstrate the efficacy of the leading-order approximation in Figure~\ref{fig:6}.

\subsubsection{Double-well potentials}
A gross deviation from OU is that of double-well potentials, and quite remarkably we find that the leading-order approximation still performs well capturing both the quantitative features and the qualitative behaviour, see Figs \ref{fig:7},\ref{fig:8}. We take a very general form 
\[
f_{Y}(\infty,y)=Ke^{-y^{2}/2}\frac{y^{2}+\gamma^{2}}{\big((y-\alpha_{1})^{2}+\beta_{1}^{2}\big)\big((y-\alpha_{2})^{2}+\beta_{2}^{2}\big)}
\]
with zeros at $y=\pm\I\gamma$ and poles at $y=\alpha_{j}\pm\I\beta_{j}$
for $j=1,2$; in \cite{Martin15b} a limited case with just the quadratic in the numerator, but no denominator, that does not allow the flexibility to explore the parameters. 
These act as follows: $\gamma\to0$ makes the
 two wells disjoint, so that it becomes progressively less easy to transit from one well to the other; $\alpha$ controls their location; $\beta\to0$
makes them deeper. The force-field is 
\[
\ffield(y)=-y+\frac{2y}{y^{2}+\gamma^{2}}-\frac{2(y-\alpha_{1})}{(y-\alpha_{1})^{2}+\beta_{1}^{2}}-\frac{2(y-\alpha_{2})}{(y-\alpha_{2})^{2}+\beta_{2}^{2}}.
\]
and there are explicit forms for $\langle-\ffield'\rangle_\infty$, and the normalising constant $K$, that can be obtained from the complex error function, or which can simply be evaluated numerically. We have evaluated several parameter sets and the results are shown in Figures~\ref{fig:7},\ref{fig:8} are typical. To be specific, the parameters are: poles at $\pm2\pm\I$, zeros at $\pm\I/\!\sqrt{2}$, and for this we have $\langle-\ffield'\rangle_\infty\approx1.557$. 
It is evident that starting from the equilibrium point (Figure~\ref{fig:7}) produces different results from starting in one of the wells (Figure~\ref{fig:8}). In the former, the approximation is excellent whilst 
 in the latter, the approximation overestimates the rate at which the process `finds out about' the other well, with the density being shared between the two wells at too early a time, though of course as $\tau\to\infty$ the results must again agree. Nonetheless even when starting in one of the wells the leading-order approximation gives qualitative insight. It is perhaps surprising that the approximation works at all, and that it does is suggestive that our philosophy of building an approximation based upon `how diffusions think about themselves' and using the slightly counter-intuitive approach of using a product, rather than a sum, expansion is of value.  

\notthis{
\begin{center}
\begin{tabular}{rrrr}
\hline
No. & Poles & Zeros & $\langle-\ffield'\rangle_\infty$ \\
\hline
1 & $\pm2\pm\I$  & $\pm\I/\!\sqrt{2}$ & 1.557 \\
2 & $\pm2\pm\I$  & $\pm\I/4$ & 2.647 \\
3 & $2\pm\I$, $-2\pm\I/\sqrt{2}$  & $\pm\I/\!\sqrt{2}$ & 1.893 \\
\hline
\end{tabular}
\end{center}

In Case~2 the wells are now considerably `more disjoint' than they are in Case~1, so that (in the language of particles in potential wells) it requires more energy to transit from one well to the other; equivalently, the process stays in one well more, or switches less.

It is evident that starting from the equilibrium point (Figure~\ref{fig:7}) produces different results from starting in one of the wells (Figure~\ref{fig:8}). In the latter, the approximation overestimates the rate at which the process `finds out about' the other well, with the density being shared between the two wells at too early a time, though of course as $\tau\to\infty$ the results must again agree. To obtain better accuracy one must take more terms in the series, as in \cite{Martin15b}. 

In the limit of disjoint wells, even the Fokker--Planck equation itself breaks down, because the process remains stuck in the well in which it started. In fact there are two steady state solutions, and ergodicity is lost because time-averaging just notices one well whereas realisation-averaging notices both. In this extreme case the methods presented here cannot work. 

*** Seems Figs 11-12 are good, Figs 9-10 less so. Comment. ***
} 

\section{Multivariate theory}
\label{sec:multi}
Buoyed by the success of the one-dimensional theory, and the exemplars that demonstrate the viability of the approach we employ across a range of illustrative cases, we move to higher dimensions where the availability of a fast accurate approximation to partially cure the 'curse of dimensionality' is attractive. In extending to the multivariate case we use two approaches: we can consider the case of the multivariate OU process as a guideline, and we can attempt to glue together components from the general one-dimensional case. There are, however, some preliminaries before we embark upon this. The general form (noting the discussion in \cite{moller10a} regarding assumptions required for the multivariate Lamperti transformation to move from $X_t$ to $Y_t$) is
\begin{equation}
dY_t = \kappa \ffield(Y_t) \, dt + \sqrt{2\kappa} \, dW_t
\label{eq:mvou_gen}
\end{equation}
where, in $m$ dimensions, $W_t, Y_t\in\R^m$ and $A:\R^m\to\R^m$ is the force field and 
 the corresponding FPE is (with $\tau=\kappa t$ as before)
\begin{equation}
\pderiv{f}{\tau} = - \nabla \cdot (Af) + \nabla^2 f.
\label{eq:mfp_f}
\end{equation}
It is natural to take $\ffield$, the force field, conservative: that is to say it is the gradient of a potential. Indeed,
\(
\ffield = \nabla \log f_\infty,
\) 
where $f_\infty$ is the steady-state solution and then $g=f/f_\infty$ obeys the adjoint equation
\begin{equation}
\pderiv{g}{\tau} = A\cdot \nabla g + \nabla^2 g
\label{eq:mfp_g}
\end{equation}
so that if $H=-g\inv \nabla g$, we have the vector equation
\begin{equation}
\pderiv{H}{\tau} = \nabla (A\cdot H + \nabla \cdot H - H \cdot H).
\label{eq:mfp_H}
\end{equation}
If $\ffield$ is non-conservative then the equations (\ref{eq:mfp_g},\ref{eq:mfp_H}) are no longer correct.

\subsection{Symmetric OU}
By a symmetric OU process, we mean (in normal form) that 
\begin{equation}
dY_t = -\kappa \ml Y_t \, dt + \sqrt{2\kappa} \, dW_t
\label{eq:mvouy}
\end{equation}
where  italic bold letters are square matrices and $W_t$ is an $m$-dimensional standard Brownian motion, i.e.\ different coordinates are independent and so $\ex[dW_t \, dW_t^\dagger]=\mI\, dt$  where $^\dagger$ denotes the transpose. The force field is conservative if, and only if, the matrix $\ml$, which we call the generator, is symmetric, and this is assumed henceforth. 
The steady-state is a Gaussian distribution of mean 0 and with covariance matrix $\msigma_\infty=\ml\inv$. Further, writing $\mq$ as the matrix exponential
\[
\mq = \exp(-2\ml \tau),
\]
which is evaluated by diagonalising $\ml$, this is eased by noting that $\ml$ is symmetric, we find the following
\begin{eqnarray}
\msigma(\tau) &=& \ml\inv(\mI-\mq) \label{eq:ousymm}\\
\fy(\infty,y) &=& |\ml/2\pi|^{1/2} \exp(-\shalf y^\dagger \ml y)   \nonumber\\
\gy(\tau, y\cdl \yzero) &=& \frac{1}{|\mI-\mq|^{1/2}} \exp \left(
-\half y^\dagger \frac{\ml \mq}{\mI-\mq} y
-\half \yzero^\dagger \frac{\ml \mq}{\mI-\mq} \yzero
+ \yzero^\dagger \frac{\ml \sqrt{\mq}}{\mI-\mq} y
 \right) \nonumber\\
&=& \frac{1}{|\mI-\mq|^{1/2}} \exp \left(
-\half (y-\yzero)^\dagger \frac{\ml \sqrt{\mq}}{\mI-\mq} (y-\yzero) \right) \nonumber\\
&& \mbox{} \times
\exp \left( \half y^\dagger \frac{\ml \sqrt{\mq}}{\mI+\sqrt{\mq}} y \right) 
\exp \left( \half \yzero^\dagger \frac{\ml \sqrt{\mq}}{\mI+\sqrt{\mq}} \yzero \right) \nonumber \\
H(\tau, y\cdl \yzero) &=& \frac{\ml \sqrt{\mq}}{\mI-\mq} (y-\yzero) + \frac{\ma\sqrt{\mq}}{\mI+\sqrt{\mq}} y \nonumber
\end{eqnarray}  
where the notation $|\cdot|$ denotes the determinant of the matrices. A technical, and notational, point is that as $\mq$ lies in the commutative matrix ring $\R[\ml]$, we can legitimately write rational functions of $\ml,\mq$ as if they were scalar indeterminates, providing the denominator is an invertible matrix. Thus $\ml/(\mI-\mq)$, $\ml(\mI-\mq)\inv$, $(\mI-\mq)\inv\ml$ are all equivalent. It is notable that the equation for $h$ is, as in one dimension \eqref{eq:OU}, a simple sum of two terms. 

\subsection{General theory}
With the OU case in hand we now proceed to the non-OU case, and emphasise that although $\ffield$ is not linear, it has to be a conservative field; broadly we follow the approach of the univariate case, but there are technicalities associated with higher dimension. Comparing with the univariate case, the next step is to define a matrix analogue of $q$ 
 and also of $\theta$. Recalling the previous discussion on Fisher information, we use the following for the multivariate analogue:
\begin{equation}
\mq = \exp(-2\mtheta \tau), \qquad {\rm and} \qquad
\mtheta = \langle -\nabla A \rangle = \langle -\nabla \nabla \log f_\infty \rangle .
\end{equation}
As $\ffield$ is conservative, its matrix $\mtheta$ of partial derivatives is symmetric and is also equal the Hessian of $-\log f_\infty$. It is then immediate that $\mtheta$ is positive definite because of the identity
\[
-\nabla \nabla (\log \psi) =  \frac{\nabla \psi}{\psi} \frac{\nabla \psi}{\psi} - \frac{\nabla \nabla \psi}{\psi};
        \]
setting $\psi=f_\infty$, multiplying by $f_\infty$ and integrating over $\R^m$ causes the second term to vanish, while the first is the integral of a tensor square.
 Analogously to the univariate case, and also (\ref{eq:ousymm}), we adopt the ansatz 
\begin{equation}
H(\tau,y) = \frac{\mtheta \!\sqrt{\mq}}{\mI-\mq} (y-\yzero) +  \frac{\sqrt{\mq}}{\mI+\!\sqrt{\mq}}  \ffield(y)  + \sqrt{\mq} \, o(1) .
\label{eq:hvec1}
\end{equation}
The first term may be integrated immediately to give a Gaussian, but the second is not in general a conservative field, except in certain cases: (i) any one-dimensional model or composite of one-dimensional models (by multiplying the marginals); (ii) any spherical model, i.e.\ one in which $f(\infty,y)$ is a function of $(y-\mu)^\dagger(y-\mu)$ for some constant vector $\mu$. This seems to be a difficulty as, no well-defined antiderivative exists in general. Upon further investigation the `non-conservative' part of this term is $O(\tau)$ because, if we write
\[
B(y) = \frac{\sqrt{\mq}}{\mI+\sqrt{\mq}}  \ffield(y) ,
\]
then
\[
\pderiv{B_i}{y_j} - \pderiv{B_j}{y_i} = O(\tau) 
\]
which at leading order is ignorable; it also disappears as $\tau\to\infty$.

We are then at liberty to define a function $\Omega(\tau,y)$ as 
\begin{equation}
\Omega(\tau,y) = \exp \int_{\muinfty}^y dx \cdot \left(\frac{\sqrt{\mq}}{\mI+\!\sqrt{\mq}}  \ffield(x) \right),
\label{eq:Omega}
\end{equation}
where $\muinfty$ is the long-term mean $\langle Y\rangle_\infty$, and the path of integration, which needs to be specified whenever the integrand is non-conservative, is a straight line. As the integral in \eqref{eq:Omega} is taken along a line, its computation does not present greater difficulties as the dimension is raised; for all the examples considered later it can be evaluated in closed form.
This is the main technical hurdle and we can now proceed in much the same way as the univariate case to find that 
\[
\gy(\tau,y) \sim (\ldots)_{\tau,\yzero} \exp\left( - \half (y-\yzero)^\dagger \frac{\mtheta \sqrt{\mq}}{\mI-\mq} (y-\yzero) \right) \cdot \frac{\Omega(\tau,\yzero)}{\Omega(\tau,y) }
\]
and, from reciprocity, as $\gy$ must be symmetric in $y,\yzero$ we must have
\[
\gy(\tau,y) = (\ldots)_{\tau} \exp\left( - \half (y-\yzero)^\dagger \frac{\mtheta \sqrt{\mq}}{\mI-\mq} (y-\yzero) \right) \cdot \frac{1}{\Omega(\tau,y) \Omega(\tau,\yzero)}
\]
so that the prefactor now depends on $\tau$ only.
As $\tau\to0$ we must have, as the density initially grows as a Gaussian, that
\[
\gy(\tau,\yzero) \sim \frac{|\mtheta/2\pi|^{1/2}}{|\mI-\mq|^{1/2} } \cdot \frac{1}{f(\infty,\yzero) } , \qquad \tau\to0,
\]
while as $\tau\to\infty$ we have $g(\tau,y)\to1$.
This demands a prefactor of the form
\[
\frac{1}{|\mI-\mq|^{1/2}} \left( \frac{|\mtheta/2\pi|}{ f(\infty,\muinfty)^2 } \right)^{\rho(\tau)} ,
\]
where $\rho(0)=\half$, and $\rho(\infty)=0$, in view of the work leading up to (\ref{eq:newg1}) we write
\begin{equation}
\rho(\tau) = \frac{1}{m}  \tr \, \frac{\sqrt{\mq}}{\mI+\sqrt{\mq}}.
\end{equation}
This gives our final results, the multivariate counterparts of \eqref{eq:newg1},\eqref{eq:newf1}, as 
\begin{equation}
\gy(\tau,y) \sim 
\frac{1}{|\mI-\mq|^{1/2}}  
 \exp\left( - \half (y-\yzero)^\dagger \frac{\mtheta \!\sqrt{\mq}}{\mI-\mq} (y-\yzero) \right) 
\frac{\left( \frac{|\mtheta/2\pi|}{ f(\infty,\muinfty)^2 } \right)^{\rho(\tau)}}{\Omega(\tau,y) \Omega(\tau,\yzero)}
\label{eq:newg2}
\end{equation}
and 
\begin{equation}
\fy(\tau,y) \sim 
\frac{1}{|\mI-\mq|^{1/2}}  
 \exp\left( - \half (y-\yzero)^\dagger \frac{\mtheta \!\sqrt{\mq}}{\mI-\mq} (y-\yzero) \right) 
\frac{\left( \frac{|\mtheta/2\pi|}{ f(\infty,\muinfty)^2 } \right)^{\rho(\tau)} \fy(\infty,y)}{\Omega(\tau,y) \Omega(\tau,\yzero)}
.
\label{eq:newf2}
\end{equation}

We consider the two special cases mentioned above. First, one-dimensional models: \eqref{eq:Omega} reduces to
\[
\Omega(\tau,y) = \left( \frac{f(\infty,y)}{f(\infty,\muinfty)} \right) ^{\frac{\scriptstyle \sqrt{q}}{\scriptstyle 1+\sqrt{q}}} ,
\]
and as $A$ is the logarithmic derivative of the invariant density we recover (\ref{eq:newg1}); $\muinfty$ does not enter the final expression. An extension is to consider a product of independent processes, so that $\nabla \ffield(y)$, and hence $\mtheta$ and $\mq$, are diagonal. This also leads to  simplifications particularly if we take the reversion speeds $(\theta_i)$ to be identical, then $\mtheta$ and $\mq$ are multiples of the identity matrix, and the multivariate approximation (\ref{eq:newf2}) becomes simply the product of the univariate approximations (\ref{eq:newf1}).
Secondly, the symmetric OU case which has $\ffield(y)=-\ml y$, $\mtheta=\ml$, $\fy(\infty,\mu_\infty)^2=|\mtheta/2\pi|$, and
\[
\Omega(\tau,y) = \exp\left( -\half y^\dagger \frac{\ml\sqrt{\mq}}{\mI+\sqrt{\mq}} y \right)
;
\]
so we end up with (\ref{eq:ousymm}), as we should. We now take illustrative examples and evaluate the leading-order approximations comparing to numerical simulation. 

\subsection{Fat-tailed and Double-well results}
Two cases of interest are a Student-t model, as a generalisation of that in section \ref{sec:student}, and a bivariate double-well model. For the bivariate Student-t model we take a general form\footnote{While $y_0$ continues to denote $y(0)$ and is a vector, $y_1$ and $y_2$ are being used to denote the components of $y$; this should not cause confusion.}
\[
\ffield(y) = -\frac{\nu+2}{\nu} \left(1 + \frac{a_1y_1^2+a_2y_2^2}{\nu} \right)\inv \begin{bmatrix} a_1y_1 \\ a_2y_2 \end{bmatrix}
\]
for which
\[
\fy(\infty,y) = \frac{\sqrt{a_1a_2}}{2\pi} \left(1 + \frac{a_1y_1^2+a_2y_2^2}{\nu} \right) ^ {-\frac{\scriptstyle \nu+2}{\scriptstyle 2}} ,\qquad
\langle -\nabla A \rangle_\infty  = \frac{\nu+2}{\nu+4} \begin{bmatrix} \sqrt{a_1/a_2} & 0 \\ 0 & \sqrt{a_2/a_1} \end{bmatrix}.
\]
Note that, writing $\mtheta$ for $\langle-\nabla \ffield \rangle_\infty$,
\[ 
\frac{\sqrt{\mq}}{\mI+\!\sqrt{\mq}} = \begin{bmatrix} \frac{\exp(-\theta_{11} \tau)}{1+\exp(-\theta_{11} \tau)} & 0 \\ 0 & \frac{\exp(-\theta_{22} \tau)}{1+\exp(-\theta_{22} \tau)} \end{bmatrix} = \begin{bmatrix} Q_1 & 0 \\ 0 & Q_2 \end{bmatrix} \quad \mbox{say},
\]
and since $\mu_{\infty}=0$ we have
\[ 
\Omega(\tau,y) = \left( \frac{f_{\infty}(y)}{f_{\infty}(0)} \right) ^{\frac{\scriptstyle a_1Q_1 y_1^2 + a_2 Q_2 y_2^2}{\scriptstyle a_1 y_1^2 + a_2 y_2^2}}
\]
so \eqref{eq:newf2} is explicit. 

This model is an extension in two dimensions of the Student-t model shown previously, so it has the same characteristics, including the fat tails. 
We take a typical example, the model with $a_1=1,a_2=3$, so that the density starts off circularly-symmetric, and ends up elliptical. Two sub-cases are displayed, one with $\yzero=(-2,2)$ (figure \ref{fig:biv-22}) and the other with $\yzero=(3,1)$ (figure \ref{fig:biv31}) as contour plots. It is pleasing to see that the features are accurately captured both quantitatively and qualitatively. 

The bivariate double-well model is particularly challenging, and very far from the simple OU model, and we take a general model  
\[
\ffield(y) = -\ma y 
+ \frac{2y}{\|y\|^2 + \gamma^2} 
- \frac{2(y-\alpha_1)}{\|y - \alpha_1\|^2+\beta_1^2} 
- \frac{2(y-\alpha_2)}{\|y - \alpha_2\|^2+\beta_2^2} 
\]
where $a_1,a_2\in\R^2$, $b_1,b_2,\gamma\in\R$ and $\ma$ is a 2-by-2 symmetric matrix.
Then
\[
\fy(\infty,y) = K e^{-y^\dagger \ma y /2} \frac{\|y\|^2+\gamma^2}{\big(\|y-\alpha_1\|^2+\beta_1^2\big)\big(\|y-\alpha_2\|^2+\beta_2^2\big)}
\]
We present cases illustrating different features:
\begin{itemize}
\item[(a)]
The first uses the following parameters
\[
\ma=\mI, \qquad \alpha_1 = \begin{bmatrix} 2 \\ 0 \end{bmatrix}, \quad
\beta_1 = 1, \qquad
\alpha_2 = \begin{bmatrix} -2 \\ 0 \end{bmatrix}, \quad
\beta_2 = 1, \qquad
\gamma = \shalf
\]
for which we compute numerically
\[
\mtheta \approx \begin{bmatrix} 1.2633 & 0 \\ 0 & 1.2774 \end{bmatrix}, \qquad 
K \approx 2.5352.
\]
Similarly to the 1D case, $\alpha$ controls the position, $\beta$ the depth and $\gamma$ makes the wells more disjoint as it goes towards $0$. Hence the wells are quite disjoint and located on the $y_1$-axis; we consider two sub-cases with different starting points.\\
In the first one, $\yzero=(0, 0.5)$ is equidistant from the two wells and the resulting field evolution is illustrated in Figure~\ref{fig:dw1mid}. The diffusion takes place at the same pace towards both wells and, even though we notice that the numerical solution converges slightly faster than the approximation, the agreement is reassuring.\\
The second one is the extreme case when it starts at $\yzero=(-1.5,0)$ corresponding to the bottom of a well and its evolution is shown in Figure~\ref{fig:dw1well}. As could be expected since the wells are well-separated, the approximation struggles to capture the medium-term behaviour.
\item[(b)]
Finally we consider
\[
\ma=\mI, \qquad \alpha_1 = \begin{bmatrix} 2 \\ 2 \end{bmatrix}, \quad
\beta_1 = 1, \qquad
\alpha_2 = \begin{bmatrix} -2 \\ -2 \end{bmatrix}, \quad
\beta_2 = 0.7, \qquad
\gamma = 1
\]
for which
\[
\mtheta \approx \begin{bmatrix} 0.9957 & -0.1990 \\ -0.1990 & 0.9957 \end{bmatrix}, \qquad 
K \approx 4.0767.
\]
In this example, due to the symmetry of $\alpha$, the wells are located on the $y_1=y_2$ line, and they are less separated because $\gamma$ is bigger. However the main difference is that $\beta_1 \neq \beta_2$ which implies that the wells have different depths. Again we present two sub-cases.\\
The first one starts equidistant from the wells with $\yzero=(-1,1)$ and is illustrated in Figure~\ref{fig:dw3mid}. Once more we observe that the rate of convergence is the same towards both wells and that the exact and leading-order solutions have a very similar behaviour even though the starting point is further away from the wells than in case (a).\\
The second sub-case starts at the bottom of a well which is represented by $\yzero=(1.3,1.3)$ and is shown in Figure~\ref{fig:dw3well}. As with case (a), the medium-term approximation diverges from the exact solution.
\end{itemize}


\section{Extensions}
\label{sec:extensions}
We have presented typical examples and demonstrated, via comparison with numerical simulation that the approximation is encouragingly capturing behaviour far from the OU case, despite being based, in some sense, around the OU. We now highlight three extensions: firstly, what happens when we big far from equilibrium, then half-line problems for which our existing theory is inconvenient, and then finally what happens if the field is non-conservative. 

\subsection{Far-field expansion}
The method is less accurate when the starting-point is a long way from equilibrium, and more concretely this means
\[
|\ffield(y)| \ll \theta |y-\mu_\infty|
\]
in one dimension.
At one level this represents the deviation from the chosen OU model. However it can also be thought of as follows: $|y-\mu_\infty|/\ffield(y)$ is, loosely, the time taken to get back to equilibrium, and if this is much larger than $1/\theta$, which is the reciprocal of the average speed of attraction, then we are in the far field---hence the above heuristic.

If we redevelop all the theory assuming no mean reversion (and there is little to be gained by staying in one dimension), we have, from (\ref{eq:mfp_H}),
\begin{equation}
H(\tau,y) \sim \frac{y-\yzero}{2\tau} + \frac{\ffield(y)}{2} + \nabla B_1(y) \tau + \cdots, 
\qquad \tau\to0,
\label{eq:hyfar}
\end{equation}
where we have replaced $b_1(y)$ by the gradient of a function $B_1$.
Equating the terms at $O(\tau^0)$ in (\ref{eq:mfp_H}) gives
\[
B_1 +  (y-\yzero)\cdot \nabla B_1  =  \frac{\|\ffield\|^2}{4} + \frac{\nabla \cdot \ffield}{2} .
\]
In one dimension, this gives
\[
B_1(y) = \frac{1}{y-\yzero} \int_{\yzero}^y \left( \frac{\ffield(z)^2}{4} + \frac{\ffield'(z)}{2} \right) dz
\]
which is the average of the integrand, and hence approximated by the average of the values at the endpoints.
\notthis{
(Despite my claim that little is gained by staying in one dimension, I am not clear how you solve for $B_1$ in higher dimension. One is to do it by the formal series
\[
B_1(y) = \sum_{n=0}^\infty \big( (\yzero-y)\cdot\nabla \big)^n   \left( \frac{\|\ffield\|^2}{4} + \frac{\nabla \cdot \ffield}{2} \right),
\]
which helps if $\ffield$ is known to be slowly-varying.)
} 
If $\ffield$ is slowly-varying then, going back to the higher-dimensional case,
\[
B_1(y) \approx {\textstyle\frac{1}{8}} \big( \|\ffield(y)\|^2 + \|\ffield(\yzero)\|^2 \big).
\]
Also, by the same arguments as in \S\ref{sec:oneD}, writing $\gy(\tau,y)=\ny(\tau)e^{-\int \hy}$,
\[
\ny(\tau) \sim \frac{e^{-\sbkt{ \|\ffield(\yzero)\|^2/4 + \nabla\cdot\ffield(\yzero)/2 } \tau}}{\sqrt{4\pi\tau}}
\]
and combining the results gives (again ignoring the $\nabla\cdot \ffield$ term)
\begin{equation}
\gy(\tau,y) \approx
\frac{e^{-\|y-\yzero\|^2/4\tau} e^{-\sbkt{ \|\ffield(y)\|^2+\|\ffield(\yzero)\|^2 } \tau/8}}{\sqrt{4\pi \tau \fy(\infty,y) \fy(\infty,\yzero)}} , \qquad \tau\to0
\label{eq:newgfar}
\end{equation}
or
\begin{equation}
\fy(\tau,y) \approx
\frac{e^{-\|y-\yzero\|^2/4\tau} e^{-\sbkt{ \|\ffield(y)\|^2+\|\ffield(\yzero)\|^2 } \tau/8}}{\sqrt{4\pi \tau }}
\left(\frac{\fy(\infty,y)}{\fy(\infty,\yzero)}\right)^{1/2}.
\label{eq:newffar}
\end{equation}

Now let us reconsider the one-dimensional dry-friction case, which we reproduce for convenience:
\[
\gy(\tau,y \cdl \yzero) = \frac{e^{-(y-\yzero)^2/4\tau}}{\sqrt{\pi \tau}} e^{-\tau/4} e^{(|\yzero|+|y|)/2}
+ \Phi\! \left( \frac{\tau - |y| - |\yzero|}{\sqrt{2\tau}} \right) .
\]
For $|y| + |\yzero|> \tau$ (a V-shaped domain, in a plot of $y$ vs $\tau$) the first term predominates, and in fact is identical to what we have just derived in (\ref{eq:newgfar}) as $\|\ffield\|=1$. This is understood as the far-field solution.
Inside the `V', i.e.\ $|y| + |\yzero|< \tau$, the second term takes over; this is the term that governs the long-time limit, and the first term decays to zero then. It is elegant that these two halves combine to give the exact solution.

The basic principle is that (\ref{eq:newgfar}) is a generic way of understanding the far-field behaviour of equations of non-OU type in the far-field, where the drift is small: for example, when $\ffield(y)=-a\tanh \gammahat y$ or $-y/(1+\gammahat^2y^2)$ as previously studied.
However, extending (\ref{eq:hyfar},\ref{eq:newgfar}) to a complete solution that works in all r\'egimes does not seem to be straightforward---there is no exponential damping factor in (\ref{eq:hyfar}), so the expression is clearly wrong as $\tau\to\infty$. This is a matter for further research.

\subsection{Square-root process}

Another new branch of the theory concerns processes that are bound to lie in the half-line by reason of the volatility decaying to zero at some point, without loss of generality $X=0$.
(This is as opposed to a reflecting boundary condition, which is another way of constraining a process; we do not consider that case here, but doubtless it could be.)
A famous example of this is the so-called square root process, described in, for example, \cite{Cox85,Lamberton12}:
\[
dX_t = a(b-X_t)\, dt + \sigma \sqrt{X_t} \, dW_t, \qquad X\ge0.
\]
It is unhelpful to transform this into (\ref{eq:sde_y}) because the drift $\ffield$ will become infinite at the origin: $1/\sigma_X$ becomes undefined.
The problem can be rescaled by setting $Y=2aX/\sigma^2$ and $\tau=at$, giving
\[
\fy(\infty,y) = \frac{y^{\nu-1}e^{-y}}{\Gamma(\nu)} , \qquad \nu=2ab/\sigma^2
\]
and
\begin{eqnarray*}
\fy(t,y \cdl \yzero) &=& 
\displaystyle
\frac{e^{-y}}{1-e^{-\tau} } \left(\frac{ye^{\tau}}{\yzero}\right)^{(\nu-1)/2}  \exp\left({-\frac{(y+\yzero)e^{-\tau}}{1-e^{-\tau}}} \right) I_{\nu-1} \left( \frac{2\sqrt{y\yzero e^{-\tau}}}{1-e^{-\tau}} \right) 
\\
\gy(t,y \cdl \yzero) &=& 
\displaystyle
\frac{\Gamma(\nu)}{1-e^{-\tau} } \left(\frac{e^{\tau}}{y \yzero}\right)^{(\nu-1)/2} \exp\left({-\frac{(y+\yzero)e^{-\tau}}{1-e^{-\tau}}} \right) I_{\nu-1} \left( \frac{2\sqrt{y\yzero e^{-\tau}}}{1-e^{-\tau}} \right) 
\end{eqnarray*}
where $I_\nu(z)$ denotes the modified Bessel function of the first kind.
We have for short time
\[
\hy(\tau,y) = \frac{e^{-\tau/2}(1-\sqrt{\yzero/y})}{1-e^{-\tau}}  - \frac{e^{-\tau/2}}{1+e^{-\tau/2}} + \frac{\nu-\half}{2y} + o(1), 
\qquad \tau\to 0
\]
where we have used $I_{\nu-1}(z)\sim e^z/\sqrt{2\pi z}$ as $z\to\infty$.
We can replace $\frac{\nu-\half}{2y}$ with $\frac{e^{-\tau/2}}{1+e^{-\tau/2}} \frac{\nu-\half}{y}$, thereby keeping an error term of the same order, $O(\tau)$, but ensuring $h(\infty,y)=0$.
This gives 
\begin{equation}
\hy(\tau,y) = \frac{e^{-\tau/2}(1-\sqrt{\yzero/y})}{1-e^{-\tau}}  + \frac{e^{-\tau/2}}{1+e^{-\tau/2}} \biggr( {-1} + \frac{\nu-\half}{y} \biggr) + o(1), 
\end{equation}
which is an analogue of (\ref{eq:hynew2}) for generalised square-root processes. The $y$-dependence is different, but there are obvious parallels, notably the $\frac{e^{-\tau/2}}{1-e^{-\tau}}$ in the first term.

The expansion can also be derived directly, if we define the normal form of a process on the half-line as
\begin{equation}
dY_t = \kappa \ffield(Y_t) \, dt + \sqrt{2\kappa Y_t} \, dW_t
\end{equation}
from which 
\begin{equation}
\pderiv{\hy}{\tau} = \pderiv{}{y} \left\{ 
\ffield(y) \,  \hy +  y\pderiv{\hy}{y} - y\hy^2  
\right\}.
\label{eq:pde_h_sqrt}
\end{equation}
In the expansion of $\hy$, the first term must, by dominant balance of the $\partial h/\partial\tau$ and $\partial(y\hy^2)/\partial y$ terms, look like
\[
\hy(\tau,y) \sim \frac{1-\sqrt{c/y}}{\tau}, \qquad \tau\to 0;
\]
the constant $c$ must equal $\yzero$, by consideration of the initial behaviour, and to ensure $h(\infty,y)=0$  we write
\[
\hy(\tau,y) \sim \frac{2\theta \!\sqrt{q} (1-\sqrt{\yzero/y}) }{1-q} + \cdots
\]
as the first term, where $\theta>0$ as before is arbitrary.
The next term is obtained, as before, by requiring that the $O(\tau\inv)$ terms balance in (\ref{eq:pde_h_sqrt}) and we find
\begin{equation}
\hy(\tau,y) \sim \frac{2\theta \!\sqrt{q} (1-\sqrt{\yzero/y}) }{1-q} + \frac{\sqrt{q}}{1+\!\sqrt{q}} \left(\frac{\ffield(y)}{y} - \frac{1}{2y}\right) + \cdots
\end{equation}
where the last term ($-1/2y$) is needed to balance the $(yh')'$ term in (\ref{eq:pde_h_sqrt}). (No such term appears in (\ref{eq:hynew}), because the relevant term in (\ref{eq:pde_h}) is $h''$, which vanishes at leading order in $\tau$.)
This gives the same result as the square-root process, for which $\ffield(y)=\nu-y$ in normal form, provided we set $\theta=\half$. 

Similar considerations apply to doubly-bounded processes e.g.
\[
dX_t = (a-bX_t) \, dt + \sigma \sqrt{1-X_t^2} \, dW_t, \qquad X \in[-1,1].
\]
Both of these are examples of processes identified by Wong \cite{Wong64}, where the usual method of solution is orthogonal expansion using the Hermite, Laguerre or Jacobi polynomials.

There is also, as intimated above, the possibility of studying processes with one or two reflecting barriers.

\subsection{Non-conservative problems}

The theory of diffusion in the presence of a non-conservative force field is considerably more difficult. Even the OU case is not straightforward, but it can be related to its symmetric case as follows.
To retain the general setting let us write the SDE as
\[
dX_t = -\kappa \ml_X \, dt + \mb_X \, dW_t
\]
with $\ml_X,\mb_X$ square matrices of dimension $m$.
Writing 
\[
Y = \sqrt{2\kappa}\, \mb_X\inv X, \qquad \ml = \kappa\inv \mb_X\inv \ml_X \mb_X,
\]
we convert the SDE to its normal form (\ref{eq:mvouy}), but $\ml$ might not be symmetric.
The invariant density is still multivariate Gaussian with zero mean but the covariance matrix $\msigma_\infty$ is not $\ml\inv$: instead it is given by the Lyapunov equation,
\[
\ml \msigma_\infty + \msigma_\infty \ml^\dagger = 2\mI
\]
which is most easily solved for $\msigma_\infty$ by writing it as a set of linear equations in its elements.
Also
\[
f(\tau,y \cdl \yzero) = \frac{1}{|2\pi \msigma(\tau)|^{1/2}} \,
\exp \left( \frac{-\big(y-\mu(\tau)\big)^\dagger \msigma(\tau)\inv \big(y-\mu(\tau)\big)}{2} \right)
\]
with
\[
\mu(\tau) = e^{-\ma \tau} \yzero, \qquad
\msigma(\tau) = 2 \int_0^\tau e^{-\ml s} e^{-\ml^\dagger s} \, ds = \msigma_\infty - e^{-\ml \tau} \msigma_\infty e^{-\ml^\dagger\tau}.
\]
From this, with $H=-\nabla \log(f/f_\infty)$ as before, we can see by direct calculation that
\begin{eqnarray*}
H(\tau,y \cdl \yzero) &=& \big(\msigma(\tau)\inv-\msigma_\infty\inv\big) y - \msigma(\tau)\inv \mu(\tau) \\
&=& \big(e^{\ml\tau}\msigma_\infty-\msigma_\infty e^{-\ml^\dagger\tau} \big)\inv (y-\yzero) + \big( e^{\ml\tau}\msigma_\infty - \msigma_\infty e^{-\ml^\dagger\tau}\big)\inv \msigma_\infty \big( e^{-\ml^\dagger\tau}- 1\big) \msigma_\infty\inv y .
\end{eqnarray*}
The first term, which is singular at $\tau=0$, corresponds to the first term in (\ref{eq:hynew},\ref{eq:hvec1}), if we accept the substitutions
\[
\frac{\theta \sqrt{q}}{1-q} \rightsquigarrow \big( e^{\ml\tau}\msigma_\infty-\msigma_\infty e^{-\ml^\dagger\tau} \big)\inv, \qquad \theta \rightsquigarrow \msigma_\infty\inv
\]
(notionally, $\sqrt{q}\rightsquigarrow e^{-\ml\tau}$ or $e^{-\ml^\dagger\tau}$),
while the second term corresponds similarly, from the substitutions
\[
\frac{\sqrt{q}}{1+\!\sqrt{q}} = \frac{\theta \sqrt{q}}{1-q} \frac{1-\!\sqrt{q}}{\theta} \rightsquigarrow
\big( e^{\ml\tau}\msigma_\infty-\msigma_\infty e^{-\ml^\dagger\tau} \big)\inv \msigma_\infty\big(1-e^{-\ml^\dagger\tau}\big)
\]
and
\[
\ffield(y) \rightsquigarrow -\msigma_\infty\inv y.
\]
Therefore, in some altered way, (\ref{eq:hynew},\ref{eq:hvec1}) carry over.

This brings us on to another matter: our work concerns approximating the FP solution using the short- and long- term behaviour. Is it possible to find two OU models with the same short- and long-term behaviour, even if the medium-term behaviour is different? The answer is yes: indeed, all generators of the form
\[
\ml = (\mI+\mtxu) \msigma_\infty\inv, \qquad \mtxu\in\mathfrak{A}
\]
where $\mathfrak{A}$ is the space of skew-symmetric matrices, give rise to the same behaviour in both limits.
Formally, define two generators $\ml$ to be equivalent (notation $\siminf$ and clearly an equivalence relation) if they give rise to the same asymptotic covariance matrix. It is easy to see that two equivalent generators have the same trace.

As an example: when $\ml = \begin{bmatrix} 1 & a \\ 0 & 1 \end{bmatrix}$ we have
\[
\msigma(t) =  
\begin{bmatrix} 1-e^{-2\tau}+\shalf a^2\big(1-(1+2\tau+2\tau^2)e^{-2\tau}\big) & -\shalf a \big(1-(1+2\tau)e^{-2\tau}\big) \\ -\shalf a \big(1-(1+2\tau)e^{-2\tau}\big) & 1-e^{-2\tau} \end{bmatrix}
\]
and
\begin{equation}
\msigma_\infty = \begin{bmatrix} 1+a^2/2 & -a/2 \\ -a/2 & 1 \end{bmatrix}, \qquad 
\msigma_\infty\inv = \frac{1}{1+a^2/4} \begin{bmatrix} 1 & a/2 \\ a/2 & 1+a^2/2 \end{bmatrix}.
\label{eq:example}
\end{equation}
So the equivalence class of $\ml$ under $\siminf$ is
\[
[\ml]_{\siminf} = 
\frac{1}{1+a^2/4} \begin{bmatrix} 1-ab/2 & a/2-b-a^2b/2 \\ b+a/2 & 1+ab/2+a^2/2 \end{bmatrix}, \qquad b\in\R
\]
which contains the following elements, as it must:
\[
\begin{bmatrix} 1 & a \\ 0 & 1 \end{bmatrix}; \quad 
\frac{1}{1+a^2/4} \begin{bmatrix} 1 & a/2 \\ a/2 & 1+a^2/2 \end{bmatrix}.
\]
All elements of $[\ml]_{\siminf}$ have trace 2. 

In summary, every OU process (or generator) is equivalent under $\siminf$ to a unique symmetric one; we obtain $\msigma_\infty$ from the Lyapunov equation and then invert it to obtain the symmetric generator.

\section{Conclusions and final remarks}
\label{sec:conclude}

We have described the solution to the Fokker--Planck equation with steady state in simple, intuitive terms and demonstrated the validity of our approach with numerical examples in a range of cases: The main results are (\ref{eq:newf1}) and its multidimensional analogue (\ref{eq:newf2}). The most striking, and potentially most useful, conclusion is that even a simple expansion without the intermediate correction terms---the `leading-order' expansion---produces acceptable results for the great majority of cases, in what may be described as the `central zone' where the process spends most of its time. Perhaps surprisingly it continues to work well even if the departure from the OU model is quite gross, such as for the double-well potentials that we consider. Given the explicit nature of the formulae we provide, they are very fast to compute by comparison with a PDE solver or Monte Carlo simulation, especially in higher dimensions. Hence we anticipate that our approach can be utilised to form the core of, say, a Kalman filter avoiding linearisation or high-dimensional computation.


We have also indicated in Section~\ref{sec:extensions} other developments that deal with extensions and important side-issues. These are: the far-field expansion, where the core result works less effectively; problems constrained to lie in the half-line or channel; and also the difficult case of when the force-field is non-conservative, i.e.\ not arising from a potential field.

On the more theoretical side there remains the open question of whether (\ref{eq:hynew}) converges in a neighbourhood of $\tau=0$, or whether it is simply an asymptotic expansion for small $\tau$ that eventually blows up if too many terms are taken. The empirical evidence of \cite{Martin15b} is that it is convergent when $\ffield$ is analytic, but this is far from clear. There is also the possibility, as indicated in \cite{Martin15b}, of using spectral methods to produce a higher-order expansion, rather than developing in a power series in $1-e^{-2\theta \tau}$ or $1-e^{-\theta \tau}$.
Another avenue is attempting to combine the far-field expansion with the near-equilibrium expansion, essentially by allowing $\theta$ to be lower when $y$ is a long way from equilibrium. If this is to be done then $\theta$ must be made symmetric in $y$ and $\yzero$ so as to preserve the reciprocity condition (\ref{eq:revers}).

This paper is, therefore, not the last word on the subject, but provides opportunities for further work in an area that we consider still to be fertile.



\notthis{

\appendix
\section{Appendix}
\subsection{Notes on integration of non-conservative fields}

If $\ffield$ is a vector field and we define
\[
\Omega(y) = \int_\xi^y dx\cdot \ffield(x)
\]
then in general:
\begin{itemize}
\item
Provided $\ffield$ is Riemann integrable, $\Omega$ is a differentiable function of $y$, but $\nabla \Omega$  \emph{does not} equal $\ffield(y)$;
\item
$\nabla \Omega$ \emph{does} depend on $\xi$.
\end{itemize}
As an example:
\[
\ffield(x) = \ml x, \qquad \Omega(y) = \int_\xi^y dx\cdot \ffield(x)
\]
where the path of integration is a straight line.
Then
\[
\Omega(y) = \frac{y^\dagger \ml y}{2} - \frac{\xi^\dagger \ml \xi}{2} + \xi^\dagger \frac{\ml^\dagger-\ml}{2} y
\]
and (with $\nabla$ denoting $\partial/\partial y$)
\[
\nabla \Omega = \frac{\ml + \ml^\dagger}{2} y +
\frac{\ml - \ml^\dagger}{2} \xi.
\]
In the symmetric case, $\ffield$ is conservative and $\nabla \Omega = \ml y$, but in general $\nabla \Omega \ne \ml y$, and $\nabla \Omega$ does depend on $\xi$.

} 


\bibliographystyle{plain}

\clearpage

\newcommand{\sizsp}{0.5}

\begin{figure}[!htbp]
\centering
\begin{tabular}{lc}
(a) & \scalebox{\sizsp}{\includegraphics{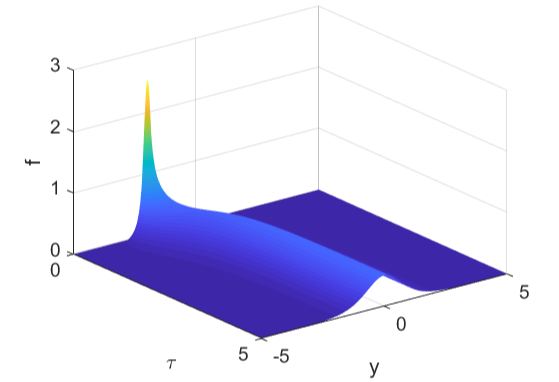}} \\
(b) & \scalebox{0.8}{\includegraphics{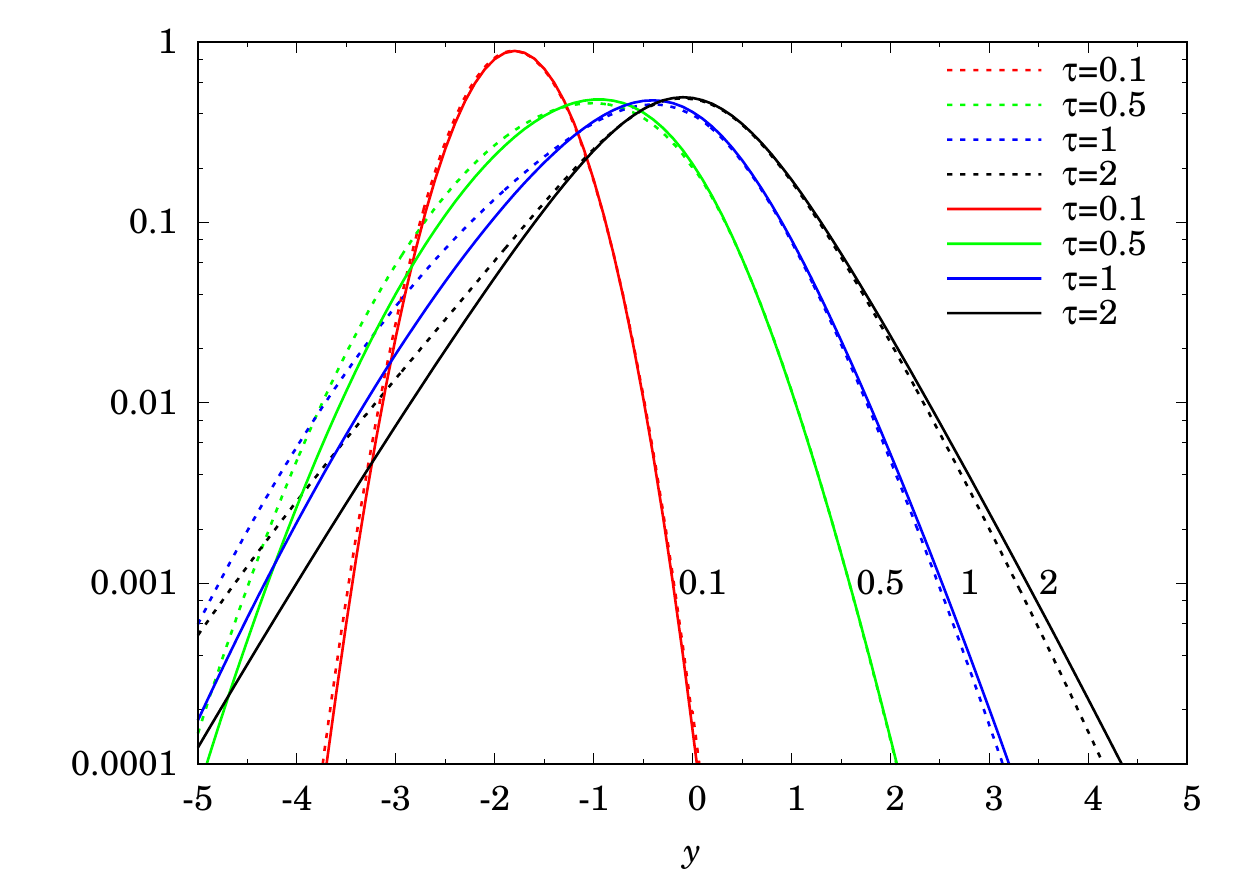}} 
\end{tabular}
\caption{
Density as a function of position ($\sech^2$, $\gammahat=1$, $\deltahat=2$), with $\yzero=-2$,
(a) the profile of $f$ versus times from the numerical simulation and (b) 
 comparing the leading-order approximation (solid) with the numerics (dotted). 
}
\label{fig:4}
\end{figure}

\begin{figure}[!htbp]
\centering
\begin{tabular}{lc}
(a) & \scalebox{0.8}{\includegraphics{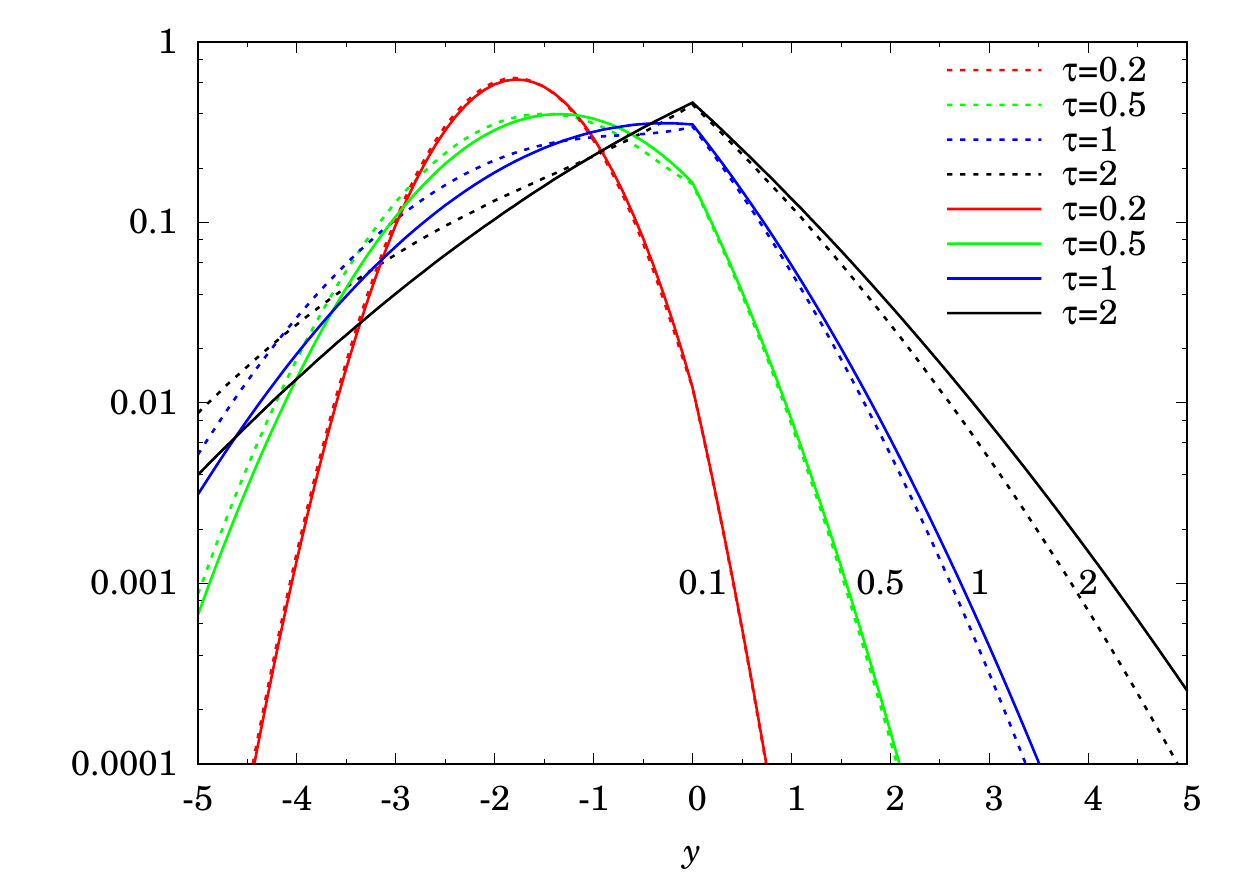}} \\
(b) & \scalebox{0.8}{\includegraphics{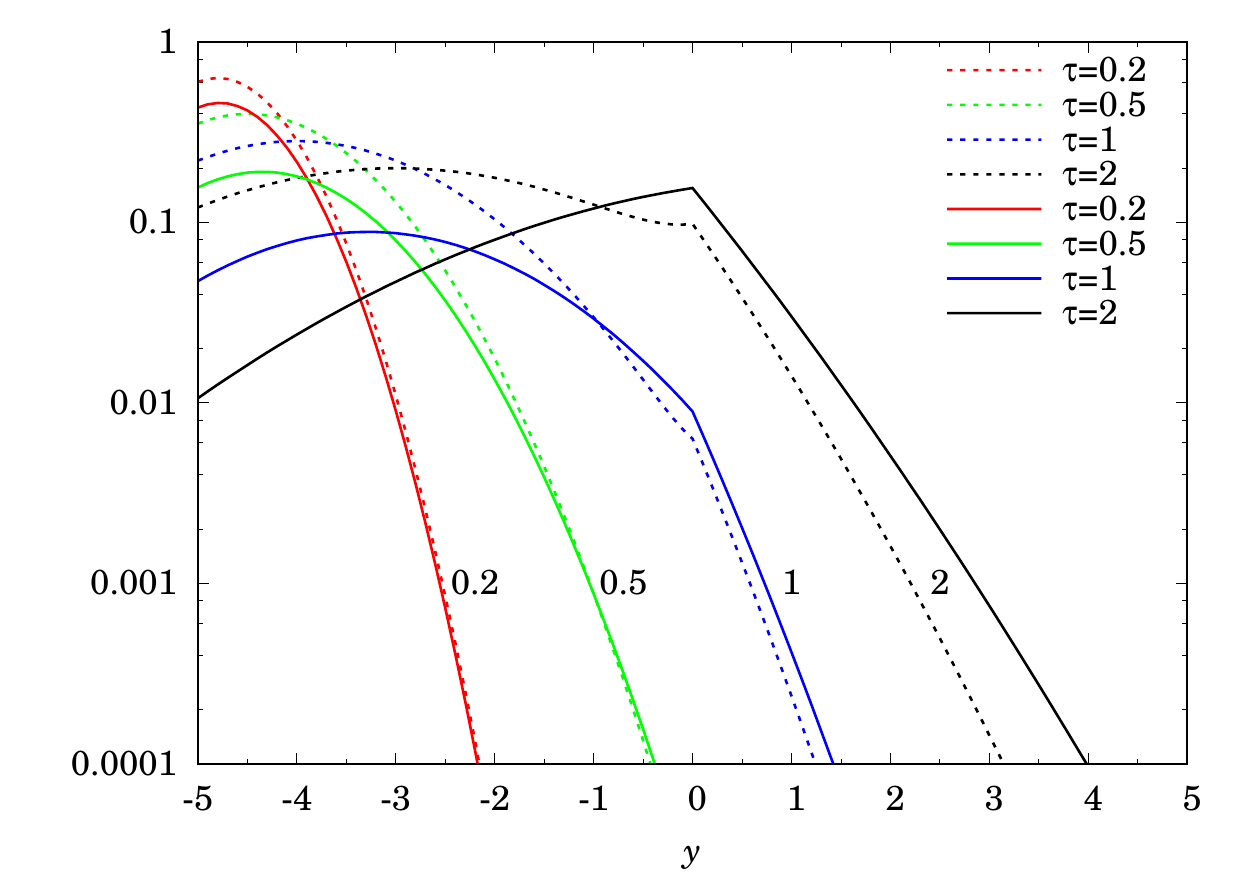}} 
\end{tabular}
\caption{
Density as a function of position (dry-friction), with (a) $\yzero=-2$, (b) $\yzero=-5$ at several different time points and comparing the leading-order approximation (solid) with the numerical simulation (dotted).  
}
\label{fig:5}
\label{fig:5A}
\end{figure}

\begin{figure}[!htbp]
\centering
\begin{tabular}{lc}
(a) & \scalebox{\sizsp}{\includegraphics{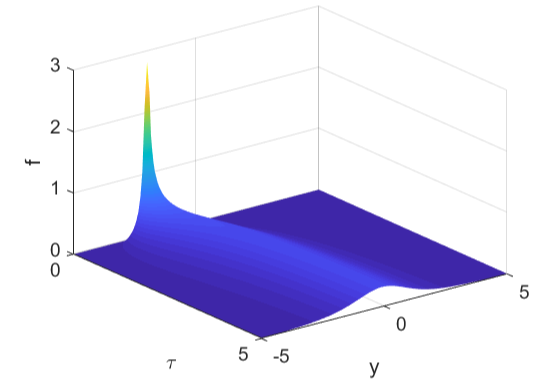}} \\
(b) & \scalebox{0.8}{\includegraphics{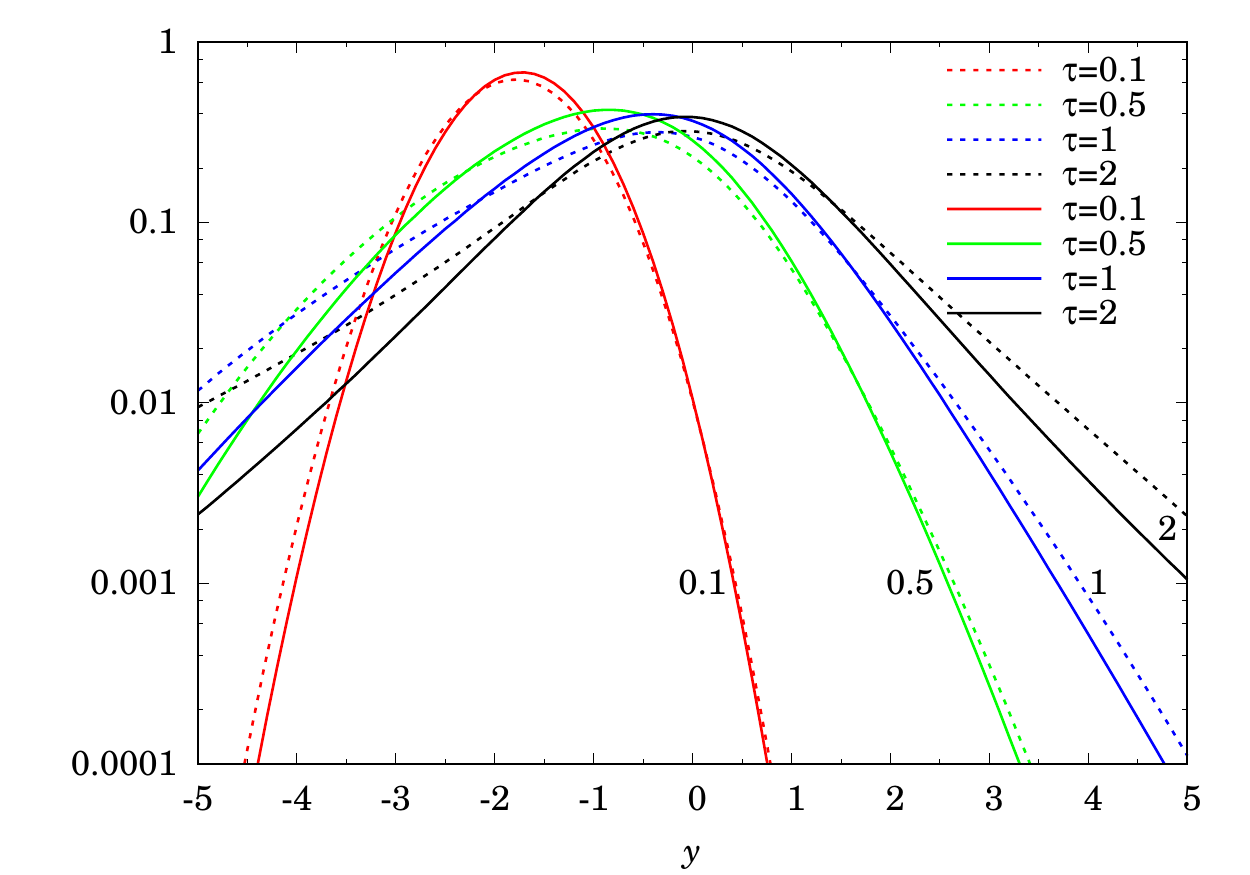}} 
\end{tabular}
\caption{
Density as a function of position (Student-t) with (a) the profile of $f$ versus times from the leading-order approximation and (b) 
 comparing the leading-order approximation (solid) with the numerics (dotted).  
}
\label{fig:6}
\end{figure}

\begin{figure}[!htbp]
\centering
\begin{tabular}{lc}
(a) & \scalebox{\sizsp}{\includegraphics{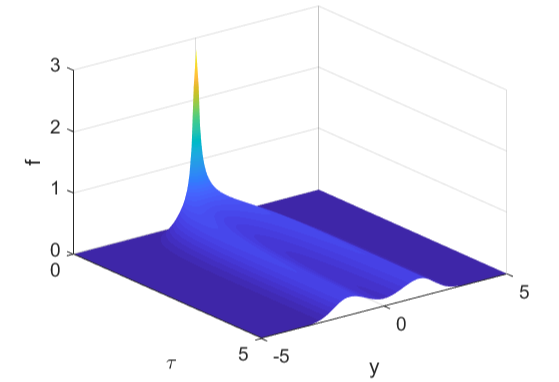}} \\
(b) & \scalebox{0.8}{\includegraphics{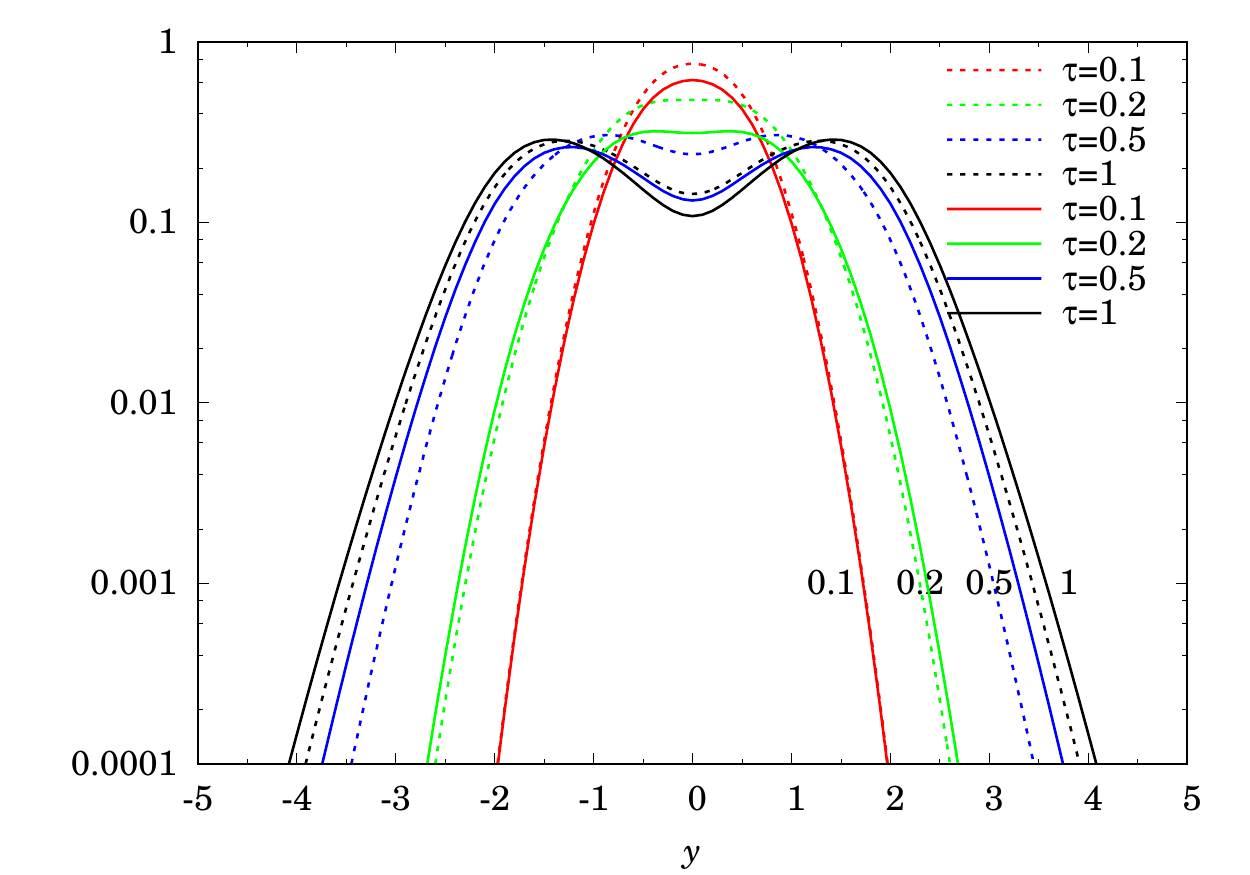}} 
\end{tabular}
\caption{
Density as a function of position (Double-well Vsn 1, poles $\pm2\pm\I$, zeros $\pm\I/\!\sqrt{2}$, $(\alpha,\beta,\gamma) = (2,1,\frac{1}{\sqrt{2}})$) with $\yzero=0$. (a) shows the field profile in time, taken from the leading-order approximation and (b) comparison of the leading-order approximation (solid) versus numerics (dotted). 
}
\label{fig:7}
\end{figure}

\begin{figure}[!htbp]
\centering
\begin{tabular}{lc}
(a) & \scalebox{\sizsp}{\includegraphics{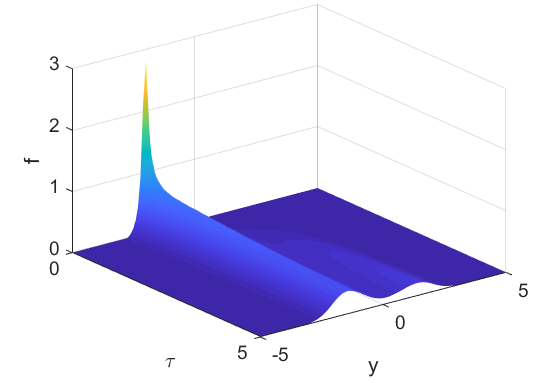}} \\
(b) & \scalebox{\sizsp}{\includegraphics{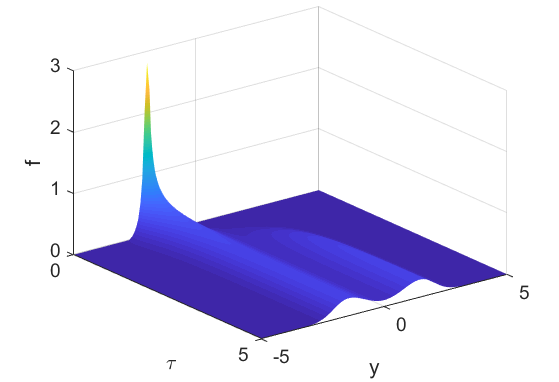}} \\
(c) & \scalebox{0.8}{\includegraphics{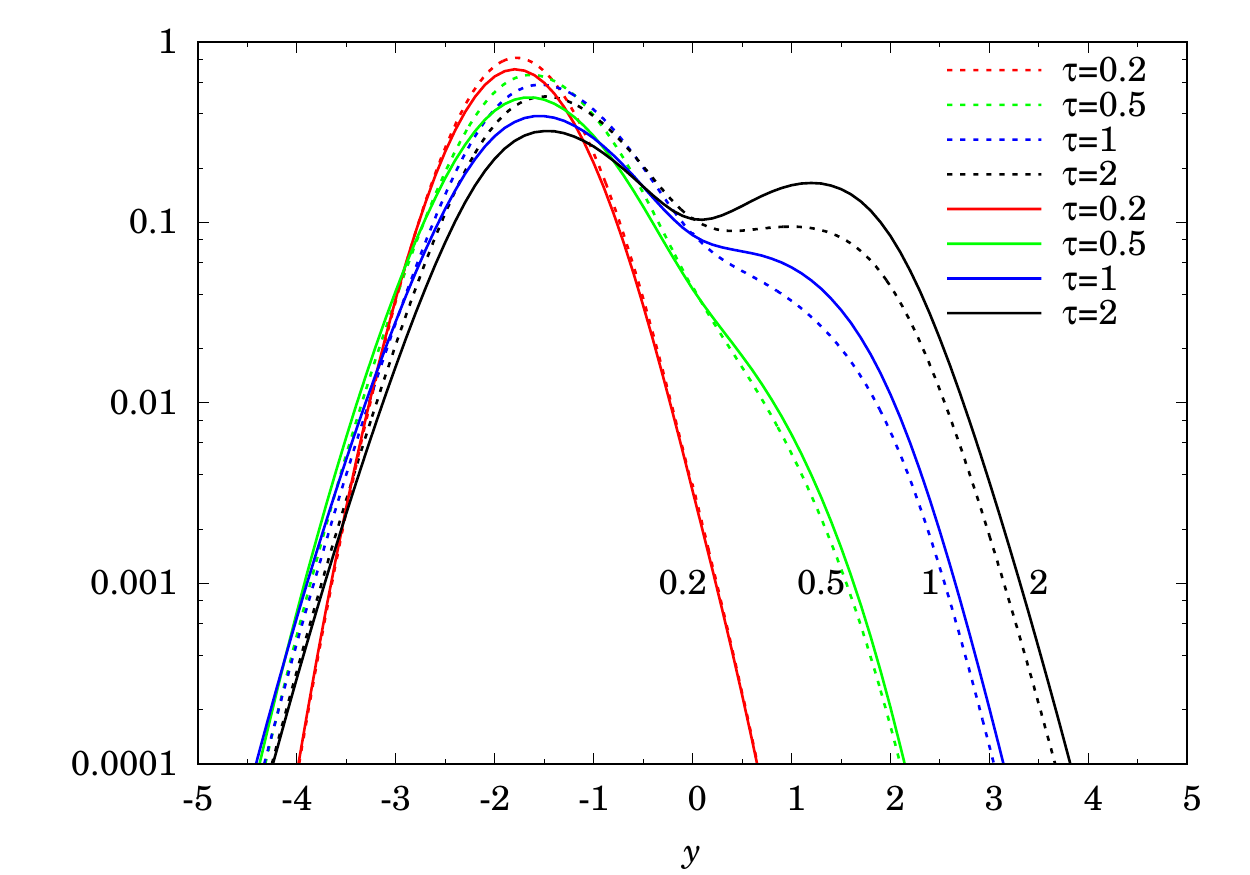}} 
\end{tabular}
\caption{
Density as a function of position (Double-well Vsn 1, poles $\pm2\pm\I$, zeros $\pm\I/\!\sqrt{2}$, $(\alpha,\beta,\gamma) = (2,1,\frac{1}{\sqrt{2}})$) with $\yzero=-2$, at several different time points:
\spiel
}
\label{fig:8}
\end{figure}

\begin{figure}[!htbp]
\centering
\begin{tabular}{lc}
\scalebox{0.9}{\includegraphics{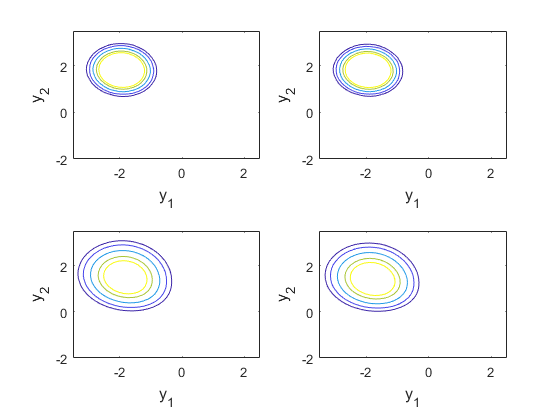}} \\
\scalebox{0.9}{\includegraphics{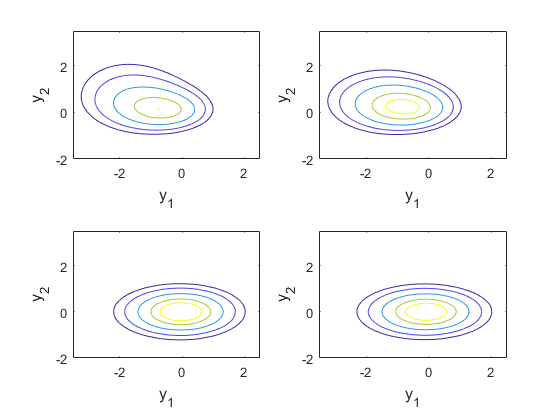}} \\
\end{tabular}
\caption{
Bivariate Student-t example (see text). PDE solver on the left, leading-order approximation on the right, $\yzero=(-2,2)$. Densities at several different time points, from top to bottom : $\tau= 0.1, 0.25, 1, 5$.
}
\label{fig:biv-22}
\end{figure}

\begin{figure}[!htbp]
\centering
\begin{tabular}{lc}
\scalebox{0.9}{\includegraphics{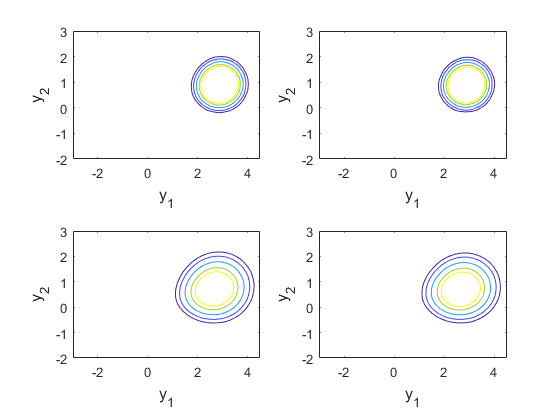}} \\
\scalebox{0.9}{\includegraphics{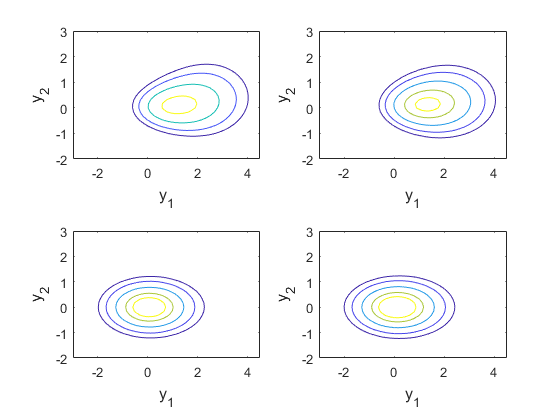}} \\
\end{tabular}
\caption{
Bivariate Student-t example (see text). PDE solver on the left, leading-order approximation on the right, $\yzero=(3,1)$. Densities at several different time points, from top to bottom : $\tau= 0.1, 0.25, 1, 5$.
}
\label{fig:biv31}
\end{figure}

\begin{figure}[!htbp]
\centering
\begin{tabular}{lc}
\scalebox{0.9}{\includegraphics{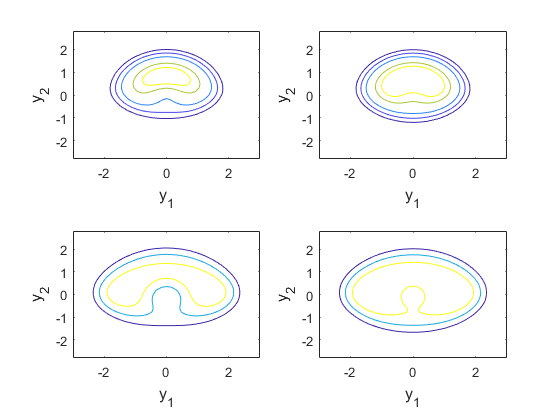}} \\
\scalebox{0.9}{\includegraphics{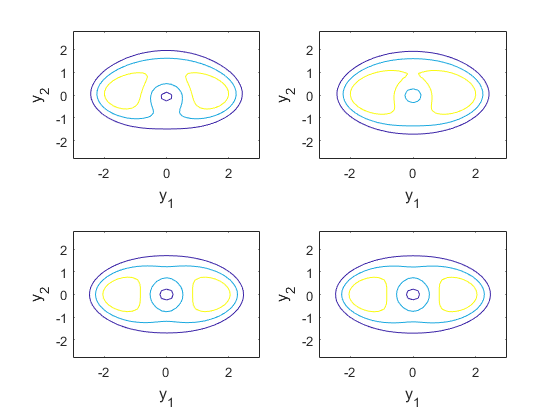}} \\
\end{tabular}
\caption{
Two dimensional double-well example (a) (see text). PDE solver on the left, leading-order approximation on the right, $\yzero=(0,0.5)$. Densities at several different time points, from top to bottom : $\tau= 0.3, 1, 1.5, 5$.
}
\label{fig:dw1mid}
\end{figure}

\begin{figure}[!htbp]
\centering
\begin{tabular}{lc}
\scalebox{0.9}{\includegraphics{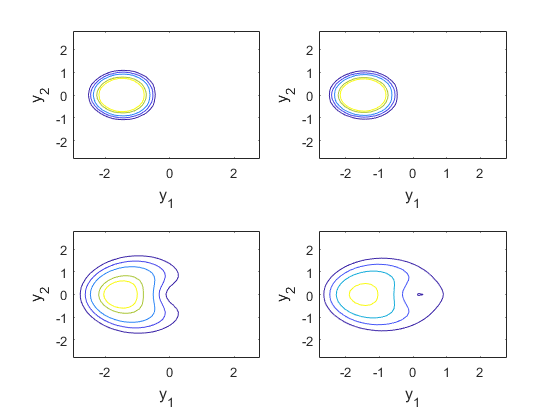}} \\
\scalebox{0.9}{\includegraphics{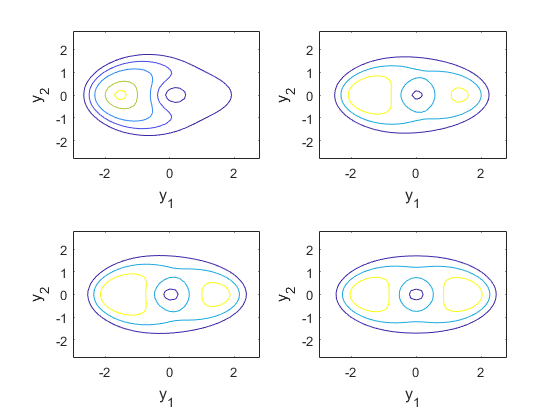}} \\
\end{tabular}
\caption{
Two dimensional double-well example (a) (see text). PDE solver on the left, leading-order approximation on the right, $\yzero=(-1.5,0)$. Densities at several different time points, from top to bottom : $\tau= 0.1, 0.8, 2, 5$.
}
\label{fig:dw1well}
\end{figure}

\begin{figure}[!htbp]
\centering
\begin{tabular}{lc}
\scalebox{0.9}{\includegraphics{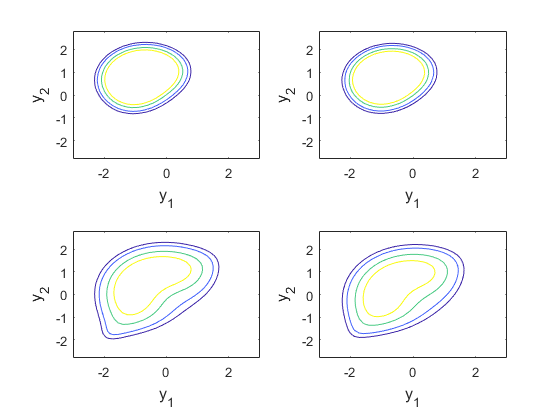}} \\
\scalebox{0.9}{\includegraphics{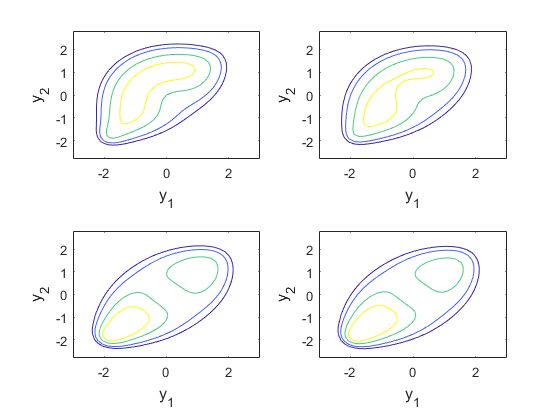}} \\
\end{tabular}
\caption{
Two dimensional double-well example (b) (see text). PDE solver on the left, leading-order approximation on the right, $\yzero=(-1,1)$. Densities at several different time points, from top to bottom : $\tau= 0.3, 1, 2, 5$.
}
\label{fig:dw3mid}
\end{figure}

\begin{figure}[!htbp]
\centering
\begin{tabular}{lc}
\scalebox{0.9}{\includegraphics{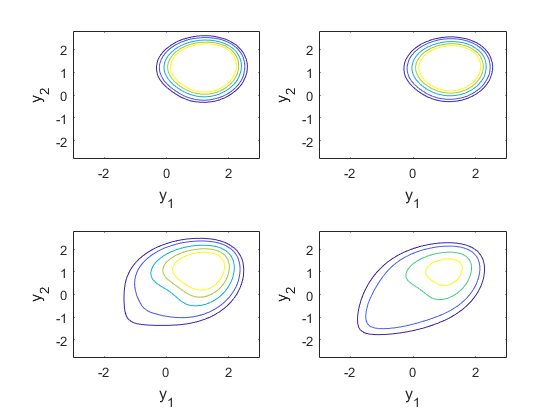}} \\
\scalebox{0.9}{\includegraphics{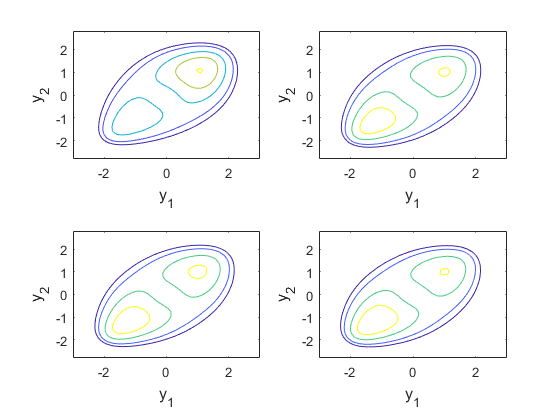}} \\
\end{tabular}
\caption{
Two dimensional double-well example (b) (see text). PDE solver on the left, leading-order approximation on the right, $\yzero=(1.3,1.3)$. Densities at several different time points, from top to bottom : $\tau= 0.3, 2, 5, 10$.
}
\label{fig:dw3well}
\end{figure}

\end{document}